\documentclass[runningheads]{llncs}
\usepackage[T1]{fontenc}
\usepackage{graphicx}
\usepackage{booktabs}
\usepackage[utf8]{inputenc}
\usepackage[frozencache,cachedir=.]{minted} 
\usepackage{makecell}
\usepackage{svg}
\usepackage{listings}
\usepackage{xcolor}
\usepackage{threeparttable}
\usepackage{pifont}
\usepackage{enumitem}
\newcommand{\cmark}{\textcolor{green!80!black}{\ding{51}}}
\newcommand{\xmark}{\textcolor{red}{\ding{55}}}

\usepackage{url}

\begin{document}
\title{Empowering Visual Artists with Tokenized Digital Assets with NFTs}
%
%
\author{Ruiqiang Li\inst{1} \and
Brian Yecies\inst{1} \and
Qin Wang\inst{2} \and 
Shiping Chen\inst{2} \and
Jun Shen\inst{1}}
\authorrunning{R.Li et al.}
%
\institute{University of Wollongong, Wollongong, NSW, Australia \and
CSIRO Data61, Sydney, NSW, Australia}
\maketitle            
\begin{abstract}
The Non-Fungible Tokens (NFTs) has the transformative impact on the visual arts industry by examining the nexus between empowering art practices and leveraging blockchain technology. First, we establish the context for this study by introducing some basic but critical technological aspects and affordances of the blockchain domain. Second, we revisit the creative practices involved in producing traditional artwork, covering various types, production processes, trading, and monetization methods. Third, we introduce and define the key fundamentals of the blockchain ecosystem, including its structure, consensus algorithms, smart contracts, and digital wallets. Fourth, we narrow the focus to NFTs, detailing their history, mechanics, lifecycle, and standards, as well as their application in the art world. In particular, we outline the key processes for minting and trading NFTs in various marketplaces and discuss the relevant market dynamics and pricing. We also consider major security concerns, such as wash trading, to underscore some of the central cybersecurity issues facing this domain. Finally, we conclude by considering future research directions, emphasizing improvements in user experience, security, and privacy. Through this innovative research overview, which includes input from creative industry and cybersecurity sdomain expertise, we offer some new insights into how NFTs can empower visual artists and reshape the wider copyright industries.

\end{abstract}
\section{Introduction}
Since 2008, blockchain technology has sparked a transformative wave of experimentation and debate, revolutionizing how data is connected to digital assets and then stored, secured, and transacted. From financial services to supply chain management, the impact of these changes has been profound and far-reaching~\cite{peters2016understanding}. The technological affordances of various blockchain developments shed light on the evolution of the wider cryptocurrency narrative. 

At the forefront of this so-called technical revolution are prominent blockchains, including Bitcoin~\cite{bonneau2015sok}, Ethereum~\cite{wood2014ethereum}, Solana (SOL), and Polkadot, as well as a motley crew of lesser-known public, private, consortium, and hybrid types. Setting the stage for this new era of digital transactions, Bitcoin, has become known as a pioneer application of blockchain technology, representing the concept of decentralized digital currency, which enables peer-to-peer transactions without the need for firmly established conventional intermediaries, such as banks. The innovation of Ethereum, on the other hand, brought smart contracts~\cite{Nick1997} to the forefront, allowing developers to create decentralized applications (dApps) and execute complex transactions in automatic ways. Cosmos~\cite{kwon2019cosmos} and Polkadot~\cite{wood2016polkadot} are two other prominent blockchain initiatives that address the scalability and interoperability challenges facing the wider blockchain space~\cite{belchior2021survey,wang2023exploring}. Cosmos is a decentralized network of independent blockchains, which enables interoperability between different blockchains through the Inter-Blockchain Communication (IBC) protocol, allowing them to communicate and transact with each other. Polkadot, founded by Dr. Gavin Wood, one of the co-founders of Ethereum, is a multi-chain network that enables blockchain interoperability, which is one of this technology's biggest challenges in terms of wide-scale adoption in society. It achieves interoperability via its unique "parachain" architecture, where multiple blockchains operate in parallel and connect to the Polkadot relay chain. Optimism~\cite{optimism} and Arbitrum~\cite{kalodner2018arbitrum} are two layer-2 scaling solutions~\cite{l2beat} designed to improve the scalability and efficiency of Ethereum-based dApps by offloading transactions from the main Ethereum blockchain~\cite{gudgeon2020sok}. Optimism utilizes Optimistic Rollups~\cite{kotzer2024sok} to increase Ethereum's transaction throughput and reduce transaction fees. Arbitrum is another layer-2 scaling solution built on Ethereum that utilizes a technology called Arbitrum Virtual Machine (AVM) and works similar to Optimism. Moreover, Solana~\cite{yakovenko2018solana} is a high-performance blockchain platform designed to support scalable dApps and cryptocurrencies with its unique consensus mechanism called Proof of History (PoH)~\cite{yakovenko2018solana}.

From a broader viewpoint, decentralized finance (DeFi)~\cite{werner2022sok,jiang2023decentralized,zetzsche2020decentralized} and non-fungible tokens (NFTs)~\cite{wang2021non} are two key innovations within the blockchain arena. Each has reshaped aspects of the digital economy in different ways. DeFi refers to a category of financial applications built on blockchain technology, aiming to disrupt traditional financial intermediaries and provide open, permission-less access to financial services. Further, DeFi protocols enable users to borrow, lend, trade, and invest in a peer-to-peer manner, without the need for traditional banks or financial institutions. These protocols typically leverage "smart contracts" on cryptocurrency platforms such as Ethereum, Polkadot, and Solana to automate financial processes, ensuring transparency, security, and efficiency. DeFi has witnessed extensive growth~\cite{jiang2023decentralized}, with projects like MakerDAO (providing decentralized stablecoins~\cite{fu2024leveraging}), Uniswap (a decentralized exchange)~\cite{adams2021uniswap,adams2023uniswap}, and Aave (a decentralized lending platform)~\cite{aave} leading to new methods of digital transactions that have begun to challenge the traditional financial sector and its long-standing, heavily central-controlled ecosystems. Due to other affordances, DeFi offers global users access to financial services previously unavailable or inaccessible, unlocking new use cases for financial inclusion and innovation. In parallel, NFTs are unique digital assets that represent ownership or proof of authenticity of a particular item~\cite{ghelani2022non}. Unlike cryptocurrencies~\cite{shirole2020cryptocurrency} such as Bitcoin or Ethereum, which on their own are fungible and can be exchanged on a one-to-one basis, NFTs are indivisible and non-interchangeable, making each one distinct and irreplaceable. They are typically created and traded on blockchain platforms, with Ethereum being a popular choice indebted to its support for smart contracts. NFTs have found applications in various domains~\cite{wang2021non}, including digital art, collectibles, gaming, and real estate, etc., extending the  shape of and processes involved in creating analog and other digital content. As a customizable digital technology, NFTs enable creators to tokenize and subsequently monetize their work by providing revenue streams~\cite{yu2024maximizing} via NFT marketplace trading platforms~\cite{ante2023non} and other Web3 environments~\cite{wang2022exploring}, which serve as alternative to traditional pathways for art trading, such as gallery exhibitions and auctions. These platforms leverage blockchain and smart contracts to offer artists, collectors, and enthusiasts innovative ways to create, showcase, trade, and engage with digital artwork. Key functionalities include artwork minting and tokenization, royalty management for content creators, curation and discovery tools, and community engagement features. Examples such as OpenSea~\cite{opensea}, SuperRare~\cite{superrare}, Foundation~\cite{foundation}, and Blur~\cite{blur2024} highlight the diverse landscape of NFT-related dApps and the process of democratizing the production, distribution, and trading of art in a Web3 landscape~\cite{wang2023exploring}. While projects such as CryptoPunks~\cite{cryptopunks}, NBA Top Shot~\cite{nbatopshot}, and CryptoKitties~\cite{cryptokitties} have gained widespread attention in conjunction of these platforms and their use of NFTs, there remains a massive community of visual artists who create unique NFT digital assets while their experiences have yet to be fully understood.

Thus far, NFTs seem poised to modify the ways artists create, distribute, and trade (i.e. monetize or 'sell') their work in the current ever-changing digital age. Traditionally, artists have faced numerous challenges in asserting ownership over their digital creations and earning a fair income from their creative productions~\cite{adler2018art}. When executed, NFTs offer an experimental solution to both aesthetic and commercial aspects of the visual arts sector by providing a unique digital certificate of authenticity and ownership, as it is stored securely on the blockchain. For different types of artists, content creators, and creative industry enterprises, NFTs represent a paradigm shift in the creative practice underpinning digital art, as well as the ownership and monetization of that work~\cite{nftdigitalartinau}. By tokenizing their work as NFTs, that is, according to how the technology works in practice, content creators can establish verifiable proof of ownership and create a kind of scarcity in the digital realm normally associated with the scarcity of physical artwork. Accordingly, this scarcity can add value to an artwork, allowing artists to sell or trade their creations as unique, one-of-a-kind digital assets~\cite{mekacher2022rarity}. 

In terms of socio-cultural dimensions, NFTs enable artists to establish more direct relationships with their audience and collectors than previously available in offline environments, bypassing traditional gatekeepers such as galleries, agents, or publishers. Through this application of blockchain technology, artwork and other digital assets can be sold and traded directly to collectors on decentralized marketplaces, eliminating intermediaries and, in some cases, retaining a higher percentage of the sale proceeds. Additionally, NFTs offer new pathways for artists to monetize their work beyond traditional sales. Through smart contracts, artists can code royalties into their NFTs, enabling them to receive a percentage of subsequent sales of their work on the secondary market~\cite{yang2023role}. This aspect of the NFT ecosystem is extremely important and empowering, as it provides artists with a potential recurring revenue stream through digital assets that are sold and traded multiple times~\cite{yu2024maximizing}. Hence, NFTs unlock some new challenges and strategies for creativity and innovation in the digital art world, including interactive, dynamic, and multimedia artwork~\cite{far2022review}. From virtual reality experiences to generative art, NFTs invite artists to extend the boundaries of creativity by experimenting with new forms of digital transactions. 

While NFTs and blockchain technologies have garnered considerable attention for their theoretical potential to transform the art industry, their actual adoption has been hindered by various security and privacy concerns~\cite{das2022understanding}. Despite their promise, these technologies are still in the early stages of development, and several challenges need to be addressed before they can achieve further adoption. Security vulnerabilities represent one of the foremost concerns surrounding NFTs and blockchain technologies. Smart contract bugs, exploits, and vulnerabilities in dApps~\cite{perez2021smart,groce2020actual} can lead to significant financial losses and reputation damage~\cite{wang2024cryptocurrency,fu2023ftx,huang2019smart}. High-profile incidents~\cite{defihacklabs}, such as the exploitation of smart contracts on DeFi platforms, have highlighted the importance of robust security measures in the blockchain ecosystem. Privacy is another critical issue facing NFTs and blockchain technologies in the visual arts. While blockchain offers transparency and immutability, it raises concerns about the protection of sensitive information stored on the blockchain~\cite{zelenyanszki2023linking}. Public blockchains like Ethereum record all transactions on a transparent ledger, potentially exposing personal data and transaction details to unauthorized parties. This lack of privacy can deter artists and collectors from fully embracing blockchain-based solutions for art management and sales. Moreover, scalability limitations on blockchain networks pose significant challenges to the adoption of NFTs~\cite{zhou2020solutions}. Nonetheless, as the demand for NFTs grows, blockchain networks may struggle to process a high volume of transactions efficiently. Network congestion and high gas fees~\cite{pierro2019influence} on platforms like Ethereum can hinder the seamless buying, selling, and trading experience, thereby discouraging participation. Interoperability issues also impede the integration of NFTs and blockchain technologies into existing art market infrastructure. Limited interoperability between different blockchain networks and platforms restricts the seamless transfer and exchange of digital assets across multiple ecosystems~\cite{belchior2021survey}. Fragmentation of this kind inhibits the development of unified and interconnected marketplaces for digital art, hindering their growth. Put simply, while NFTs and blockchain technologies hold incredible potential to transform the creative and other industries, security, privacy, scalability, and interoperability challenges remain huge barriers for their adoption.

\begin{figure}[!htp]
    \centering
    \includegraphics[width=\linewidth]{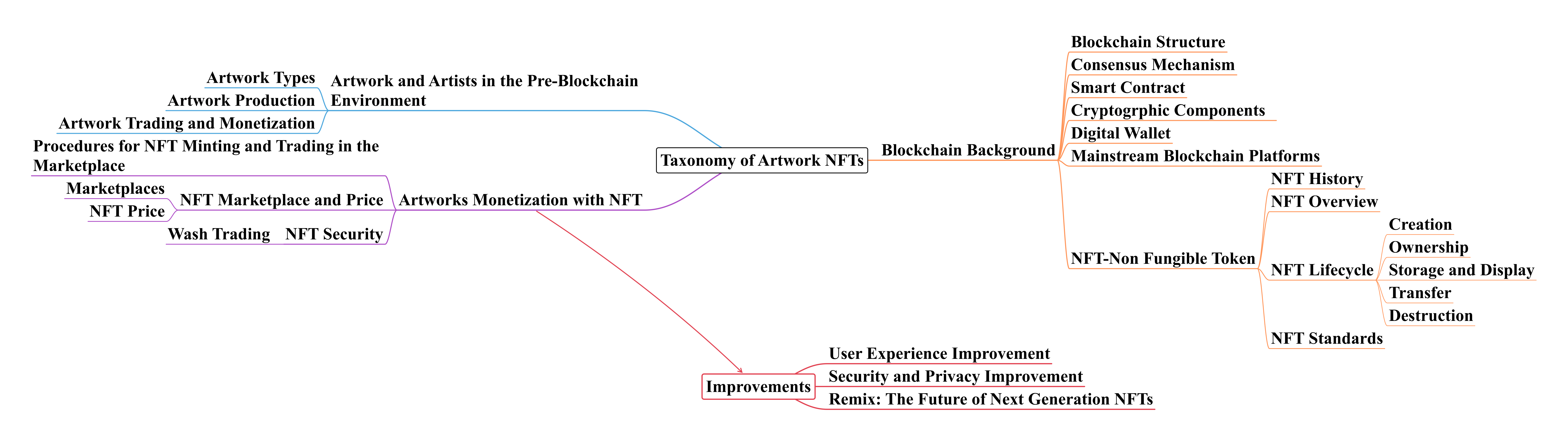}
    \vspace{-0.3in}
    \caption{Our Taxonomy}
    \label{fig:Our Taxonomy}
    \vspace{-0.4in}
\end{figure}

\noindent\textbf{Contributions.} To address some of the critical strengths and weaknesses of this emergent digital domain, in this paper, we provide a systematical overview of applying NFT in the visual arts field (cf. Figure~\ref{fig:Our Taxonomy}). The paper begins by examining artwork types based on their medium and then explores the production processes involved in creating these works in the pre-blockchain environment. It looks into how artists benefit from and monetize their traditional art through established methods. Following this, the paper introduces the basics of blockchain technology, including its structure and various consensus mechanisms of the protocols mentioned above. Smart contracts are highlighted as a crucial element in the blockchain paradigm due to their capability to automatically execute specific contractual conditions. Digital wallets are then discussed as the chief gateway for users to access the blockchain world, and an overview of mainstream blockchains is provided. Next, the paper reviews the background of NFTs, including their history, technical overview, lifecycle, and standards relevant to traditional types of artwork discussed below. With this foundational knowledge, the paper focus on investigating how the NFT ecosystem can be applied to traditional artwork, exploring how NFTs can enhance the benefits and monetization through NFT trading procedures. It also surveys existing NFT marketplaces that facilitate the trading of artwork and identifies factors that influence NFT prices. Security issues, which are critical for safe NFT trading, are also examined. Finally, the paper proposes three improvements to enhance the current NFT ecosystem: improving user experience, enhancing security and privacy, and supporting artwork remixing. Combined, these three developments aim to better empower artists and serve all stakeholders within the NFT ecosystem.

\vspace{-0.1in}

\section{Artwork and Artists Before Blockchain}
\subsection{Artwork Types} 
Artwork can be categorized in numerous ways, offering multiple windows through which to understand artistic expression. One of the most relevant and detailed categories is the medium of creation. This category focuses on the materials and techniques used in the production of an artwork, which can influence its appearance, texture, and emotional impact. Medium-based categorization includes various forms such as painting, drawing, sculpture, printmaking, photography, and certainly, the digital art (made by computing services, while not NFT-based).

Paintings, for instance, can be rendered in oils, watercolors, acrylics, or tempera, each bringing distinct qualities and challenges. Oil paintings are renowned for their rich textures and depth, allowing artists to build up layers and create detailed, vibrant compositions~\cite{elkins2019painting}. Watercolors, are celebrated for their fluidity and translucence, often used to capture delicate light and atmosphere~\cite{curtis1997computer}. Acrylics are versatile and fast-drying, making them ideal for modern and abstract works, while tempera, with its quick-drying and durable properties, has been used for centuries in religious and iconographic art. Sculpture is another prominent medium, encompassing a variety of materials such as stone, metal, wood, and clay~\cite{penny1993materials}. Stone sculptures, like those from ancient Greece and Rome, are lauded for their permanence and classical beauty. Metal sculptures, often seen in contemporary art, allow for bold, dynamic forms and intricate details. Wood offers a warm, organic quality, while clay provides flexibility and is often used for both functional pottery and expressive, detailed figurative work. Printmaking includes techniques such as lithography, etching, and screen printing, each allowing artists to produce multiple copies of a single image, expanding the accessibility of their work~\cite{abidin2013printmaking}. On the other hand, photography and digital art represent more modern mediums. Photography captures moments in time with precision and can be manipulated in darkrooms or digitally to create a wide range of effects~\cite{elkins2011photography}. Digital art, including everything from digital painting to 3D modeling, leverages technology to push the boundaries of  realism~\cite{paul2023digital}.

In addition to medium, artwork is often categorized by style or movement. This includes broad aesthetic and philosophical movements such as Renaissance~\cite{matthias_2024}, Baroque~\cite{martin2018baroque}, Impressionism~\cite{ferguson1982defining}, Expressionism~\cite{weisstein1967expressionism}, Surrealism~\cite{breton1969manifestoes}, Cubism~\cite{gantefuhrer2004cubism}, Abstract Art~\cite{pippin2002abstract}, Pop Art~\cite{madoff1997pop}, and Contemporary Art~\cite{smith2009contemporary}. Each of these styles is characterized by specific techniques, themes, and historical contexts, reflecting the evolving psychology of artistic expression. Historical period is another significant category, spanning from prehistoric art~\cite{bahn1998cambridge} to ancient civilizations like Egyptian, Greek, and Roman~\cite{winckelmann1880history}, through to Medieval~\cite{nees2002early}, Renaissance, Baroque, and into modern and contemporary times. These periods reflect the changing cultural, social, and political landscapes throughout history. Subject matter also plays a crucial role in categorizing art, with common themes including portraits, landscapes, still life, abstract compositions, historical scenes, religious motifs, and mythological tales~\cite{klavans2014subject,male1982religious,panofsky1933classical}. Each subject matter can provide insights into the artist's intentions and the cultural context of the work. Lastly, regional or cultural distinctions offer another layer of understanding. Western art, Eastern art (including Chinese, Japanese, and Indian), African art, and Indigenous art each encompass unique traditions, techniques, and narratives, reflecting the diverse heritage of human creativity, differentiated by geographic regions though modern art tends to be globalised.  

\subsection{Artwork Production} 
Artwork production has a series of steps and processes that artists normally follow to create their works. The production process can be signiﬁcantly different depending on the medium and the individual artist’s techniques and style.

The creation of artwork begins with inspiration, which can come from nature, literature, other artworks, and personal experiences. After ﬁnding inspiration, artists need to study the theme, style, techniques, and historical background related to their artwork. The next stage is planning and sketching drafts, where artists create preliminary sketches and plan the composition, color scheme, and overall design of their artwork. This phase is crucial for the ﬁnal artwork because it allows artists to experiment with different elements and make adjustments before the ﬁnal creation. Finally, the production process varies depending on the medium used, as each medium requires its own speciﬁc techniques and approaches. Table~\ref{tab:Medium Specific Processes} shows the medium specific processes. 

\vspace{-0.1in}
\begin{table}[htbp]
\caption{Artwork Production Processes}
\begin{center}

\resizebox{\linewidth}{!}{
\begin{tabular}{c|c c c c c}
  
  \textbf{ Medium }  & \textbf{ Preparation} & \textbf{ Creation} & \textbf{ Refinement } & \textbf{ Finishing } \\
  \midrule
    Painting & Surface Preparation & Underpainting & Layering/Detailing & Varnishing \\
    Sculpture & Modeling/Material Selection & Carving or Casting & Assembly & surface treatments \\
    Printmaking & Design/Plate Preparation & Inking/Printing &  & Editioning \\
    Photography & Conceptualization  & Shooting  & Editing & Printing \\
    Digital Art & Software Selection & Digital Sketching  & Layering/Rendering  & Finalization \\
\end{tabular}
}
\label{tab:Medium Specific Processes}
\end{center}
\end{table}
\vspace{-0.3in}

The ﬁnal stage is getting the artwork ready for display, which may include framing, mounting, or employing other presentation methods to showcase it effectively. Following this, its important to document details about the artwork, such as the title, creation date, materials used, and any relevant notes or context. This documentation provides valuable insight into the artwork and is crucial for potential buyers and viewers. Finally, the artwork is displayed in different venues, such as galleries, museums, online platforms, or other public spaces. At the same time, the artwork is available for sale during this stage.

\subsection{Artwork Trading and Monetization} 
Typically, trading and monetizing artwork involves various methods, platforms, and strategies that artists and collectors use to buy, sell, and earn revenue from this work. Prior to the advent of NFTs, the art trading industry had already been considered as a multifaceted industry, incorporating certain methods to connect artists with collectors. Historically, galleries have been the cornerstone of the art market, providing artists with representation, organizing exhibitions, and taking commissions from sales~\cite{reutter2001artists}. Art fairs further expand these opportunities, allowing galleries and independent artists to rent booths, display their works, and network with a broad audience of collectors, curators, and peers~\cite{thompson2011art}. Auction houses like Sotheby’s~\cite{sotheby} and Christie’s~\cite{christies} offer high-profile avenues for the sale of artwork through a competitive bidding process, often achieving significant price escalations for coveted pieces~\cite{brodie2014auction}. Private sales also play a crucial role in this off-line domain, enabling direct transactions between artists, collectors, and through art advisors who facilitate these exchanges discreetly.

Online platforms are facilitating new opportunities for artwork trading. E-commerce sites such as Etsy~\cite{etsy}, Saatchi Art~\cite{saatchiart}, and Artfinder enable artists to sell directly to consumers, while digital galleries provide virtual spaces for exhibitions. Social media platforms, such as Instagram~\cite{instagram}, Facebook~\cite{facebook}, TikTok~\cite{tikTok}, and Canva~\cite{canva}, have become powerful tools for artists to showcase their work. They engage with and make sales to a global audience in an online space. Until now, this combination of conventional methods has helped artists to reach potential buyers while enabling collectors to discover new talent and work with somewhat ease.

In the above-mentioned pre-blockchain environment, the monetization of artwork encompasses a variety of strategies that assist artists to generate income from their creations via face-to-face and online sales. Limited editions of analog work, for example, allow artists to produce a finite number of prints maintaining an element of exclusivity~\cite{bennett2013attractiveness}. Each print is often accompanied by a certificate of authenticity to verify its edition number and originality. Licensing offers another lucrative pathway, where artists grant rights to reproduce their work across a range of products such as apparel, home decor, and stationery, thereby earning royalties from these sales~\cite{hick2019artistic}. Commissions provide artists with opportunities to create custom work tailored to the preferences of individual clients or for large-scale projects in corporate and public spaces~\cite{nelson2008patron}. Merchandising further expands an artist's revenue streams, with products like t-shirts, mugs, and posters featuring their designs sold through physical market stalls and online shops~\cite{mckelvey2015merchandising}. Crowdfunding platforms like Kickstarter~\cite{kickstarter} and Patreon~\cite{patreon} allow artists to generate funding for projects through direct support from patrons, often offering exclusive content and perks to subscribers who commit to regular contributions~\cite{dalla2021crowdfunding}. Additionally, artists can share their expertise and connect with a broader audience by offering workshops and classes through online and in-person, by platforms like Skillshare~\cite{skillshare} and Udemy~\cite{udemy}. These educational courses not only exalt income but also help to build a community of followers and supporters. Together, these monetization strategies provide artists with multiple channels to sustain their creative practices and reach a wider audience. Having said this, we assert the addition of blockchain technology to any one of the above mentioned tactics offers practitioners, enterprises, and dealers new opportunities for achieving their immediate and long-term goals. 

What follows next is a technical discussion of highly relevant elements of the blockchain ecosystem, which represents the minutia of this evolving space.

\section{Blockchain}
\subsection{Blockchain Structure}
Blockchain architecture is a decentralized system that enables the secure and transparent recording of transactions across a network of computers. A blockchain architecture includes a series of components as follows.

\begin{itemize}
\setlength{\itemsep}{0pt}

\item Nodes: Individual computers or devices which are connected to the blockchain network are nodes. Each node stores a copy of the entire blockchain ledger.

\item Blocks: Transactions are grouped together into blocks, which are then added to the blockchain in a sequential and immutable fashion. Each block typically contains a cryptographic hash of the previous block, creating a chain of blocks, hence interpreting the term \textit{blockchain}.

\item Consensus mechanism: This is a type of protocol or algorithm used to achieve agreement among nodes on the validity of transactions and also the order in which they are added to the blockchain. Common consensus mechanisms typically include Proof of Work (PoW), Proof of Stake (PoS), Practical Byzantine Fault Tolerance (PBFT)~\cite{castro1999practical}, and etc.

\item Smart contracts: These are self-executing contracts with the terms of the agreement directly written into code. Smart contracts automatically execute the agreed terms when predefined conditions are met~\cite{kosba2016hawk}.

\item Cryptographic hashing: Hash functions are used extensively in blockchain to ensure the integrity and security of data. Each block contains a cryptographic hash of the previous block to resist tamper. 

\end{itemize}

\begin{figure}[!htp]
    \centering
    \includegraphics[width=0.93\linewidth]{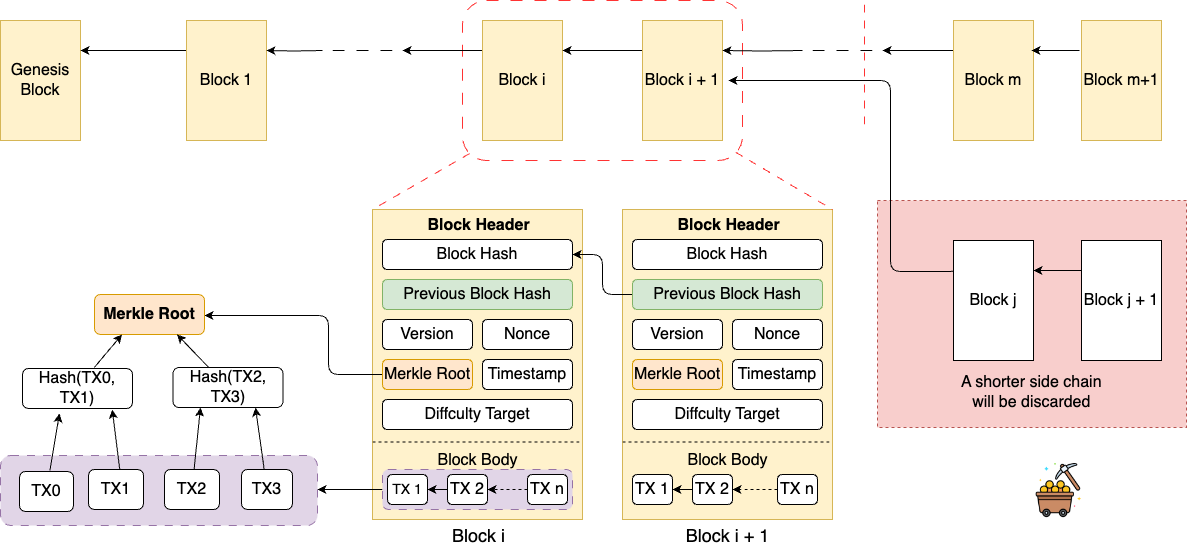}
    \vspace{-0.2in}
    \caption{Block Structure}
    \label{fig:Block Structure}
\end{figure}

Blocks are linked together in a linear sequence. This chronological approach forms a chain. The linking mechanism is achieved through referencing the hash of the previous block in each subsequent block's header. This creates a continuous chain of blocks, with each block containing the hash of the block that came prior to it. This ensures that any attempt to modify a previous block's data would invalidate all subsequent blocks in the chain. Since each block's hash is based on the data of the block itself and the hash of the previous block, even a small change in any block would cause a ripple effect. This linking mechanism ensures the immutability and integrity of the blockchain. Once a block is added to the chain, it becomes extremely difficult to modify any of its data without being detected by the network, which making blockchain a secure and reliable method for recording and verifying transactions. A blockchain typically consists of three key components, namely,  block header, transactions, and block hash. Figure~\ref{fig:Block Structure} shows the whole details for the block structure~\cite{zheng2020overview}.

\vspace{-0.1in}
\vspace{-0.2in}

\begin{itemize}
\setlength{\itemsep}{0pt}

\item 
Block header: This contains metadata about the block, including information such as the block's unique identifier (hash), timestamp of when the block was created, a nonce (a number used in mining to find a valid hash), and a reference to the previous block's hash.

\item
Transactions: These are the actual data that the block contains. Transactions include various types of data depending on the blockchain's purpose of use. In general, transactions would include sender and recipient addresses, transaction amounts, and digital signatures for verification.

\item
Block hash: This is a unique cryptographic hash~\cite{preneel1994cryptographic} generated by running the block's contents via a hashing algorithm. The hash serves as the block identifier and is crucial for ensuring the integrity of data.
\end{itemize}

\subsection{Consensus Mechanism}
There are many consensus algorithms for blockchains, and every consensus algorithm has its advantages and trade-offs, making them suitable for different use cases and blockchain implementations. The most well known consensus algorithms are PoW, PoS, Delegated Proof of Stake (DPoS),PBFT and RAFT~\cite{mingxiao2017review}. Table~\ref{tab:Consensus Algorithms Properties} shows different Consensus Algorithms' properties~\cite{zheng2017overview,mingxiao2017review}.

\vspace{-0.2in}
\begin{table}[!htbp]
\caption{Consensus Algorithms}
\vspace{-0.3in}
\begin{center}
\resizebox{1\linewidth}{!}{
\begin{tabular}{c|c c c c c}
  \textbf{  Consensus Algorithm} &\textbf{ PoW } & \textbf{ PoS } & \textbf{ DPoS} & \textbf{ PBFT} & \textbf{ RAFT } \\ 
    \midrule
    Crash Fault Tolerance & 50\% & 50\% & 50\% & 33\% & 50\% \\
    Verification speed & >100s & <100s & <100s & <10s & <10s \\
    Throughput (TPS) & <100 & <1000 & <1000 & <2000 & >10K \\
    Scalability & Strong & Strong & Strong & Weak & Weak \\
    Membership & Open & Open & Open & Permissioned & Open \\
    Energy Saving & \xmark & Partial & Partial & \cmark & \cmark \\
    Examples & Bitcoin & Ethereum 2.0 & Bitshares~\cite{bitshares} & Hyperledger Fabric~\cite{hyperledgerfabric} & Quorum~\cite{quorum} \\
\end{tabular}
}
\label{tab:Consensus Algorithms Properties}
\end{center}
\end{table}
\vspace{-0.4in}

\begin{itemize}
\setlength{\itemsep}{0pt}

\item 
PoW: PoW is the original consensus algorithm used in blockchain networks like Bitcoin. In PoW, miners compete to solve complex mathematical puzzles to validate transactions and create new blocks. The first miner to solve the puzzle broadcasts the solution to the network, and if validated, the block is added to the blockchain. PoW requires a significant amount of computational resources, making it secure but also energy-intensive.

\item 
PoS: PoS selects validators to create new blocks based on the amount of tokens that they are willing to stake. Validators are chosen pseudo-randomly (e.g., via RANDAO), often higher chances are given if they stake more cryptocurrency. PoS is considered more energy-efficient than PoW since it avoids massive computation. Ethereum has transitioned to PoS with its upgrade.

\item 
DPoS: DPoS is a variation of PoS where token holders vote for a limited number of delegates to validate transactions and produce blocks on their behalf. These delegates are usually known as \textit{witnesses} or \textit{delegated nodes}. DPoS is designed to improve scalability and efficiency by reducing the number of participants involved in the consensus process. For example, EOS~\cite{io2017eos} and TRON~\cite{tron} are using DPoS.

\item
PBFT: PBFT is designed to tolerate Byzantine faults/corruptions, where a malicious node in a network may fail or behave maliciously. PBFT works in a way by having a leader node to propose a block, which is then validated by two-thirds of the nodes. If an enough number of nodes agree on the validity of the block, it is added as the newest block in the blockchain. PBFT’s valuable properties are its high throughput and low latency, making it suitable for permissioned blockchain networks~\cite{de2018pbft}.

\item
RAFT: RAFT is designed for fault-tolerant distributed systems. It works by electing a leader among the nodes, which is responsible for producing the new block. RAFT achieves consensus consistency by a leader-based approach, where the leader coordinates the agreement among the validator nodes. RAFT has higher consensus consistency than PBFT.

\end{itemize}

\subsection{Smart Contract}

Smart contracts are central to blockchain technology because they bring programming capabilities. These self-executing digital contracts have terms written in code and are stored on a blockchain network. Smart contracts represent a transformative innovation, offering a decentralized solution to traditional contractual agreements. Their integration into blockchain networks introduces new attributes such as trust and efficiency, which can drive the application of blockchain across various industries and use cases. These smart contracts operate within the framework of blockchain with its inherent characteristics, such as immutability, transparency, and decentralization, utilizing these characteristics to ensure the integrity and security of transactions~\cite{nugent2016improving,li2022smart}.

Smart contracts are validated through consensus mechanisms, where network nodes collectively agree on the validity of transactions and the execution of contract code. This consensus ensures tamper-proof execution, because each transaction is validated and recorded across all nodes in the network.

One of the most notable blockchains for deploying smart contracts is Ethereum, which introduced the concept of a Turing-complete blockchain with its Ethereum Virtual Machine (EVM)~\cite{wood2014ethereum}. The EVM enables developers to deploy and interact with smart contracts written in languages like Solidity and Vyper, which allows for the creation of a wide range of dApps on the Ethereum platform.

However, the adoption of smart contracts on blockchains faces several challenges, including scalability, interoperability, legal compliance, and security vulnerabilities. These issues require careful consideration~\cite{zheng2020overview}. Despite these challenges, the potential of smart contracts continues to motivate researchers, developers, and industry leaders to explore new possibilities and try to expand the boundaries of decentralized technology~\cite{tolmach2021survey,wan2021smart,li2022sok}.

\subsection{Cryptogrphic Components}

In blockchain technology, hash and signature algorithms are crucial for maintaining security and integrity. SHA-256~\cite{gilbert2003security} is a widely used hash algorithm in blockchains, generating a 256-bit hash for each block. This hash ensures consistency and immutability~\cite{thuy2020fast}. In Ethereum, Ethash algorithm~\cite{ethash} is used as the PoW consensus algorithm. Ethash combines SHA-3~\cite{dworkin2015sha} with a memory-hard design, increasing mining difficulty and preventing centralization. Additionally, Ethereum employs Keccak-256 (SHA-3) to compute the hash for each block, enhancing security. For transaction verification, both Bitcoin and Ethereum use digital signatures. The most common signature algorithm is Elliptic Curve Digital Signature Algorithm (ECDSA)~\cite{mayer2016ecdsa}, which employs elliptic curve cryptography to verify the authenticity and integrity of transactions. These hashing and signature algorithms collectively help blockchains ensure data integrity, transaction authenticity, and resistance to tampering.

\subsection{Digital Wallet}

A blockchain wallet is the entrance for users to manage, store, operate with the cryptocurrencies and digital assets securely~\cite{chatzigiannis2022sok,suratkar2020cryptocurrency}. Unlike traditional wallets, which hold physical cash or cards, blockchain wallets are also digital interfaces that allow users to access their funds on blockchain networks. A wallet consists of two components: a public address and a private key. The public address is the destination for receiving funds and can be shared openly with others. The public address has similar function to a bank account number in traditional finance. On the other hand, the private key is a secret code that controls access to the wallet and allows the owner to authorize transactions. The private key act as the password to unlock and manage the funds stored in the wallet. 

\vspace{-0.2in}
\begin{table}[htbp]
\caption{Wallet Security}\label{tab1}
\vspace{-0.2in}
\begin{center}
\resizebox{0.8\linewidth}{!}{
\begin{tabular}{c|c c c c c}
  \textbf{   Wallet Types } & \textbf{  Software} & \textbf{ Hardware} & \textbf{ Paper} & \textbf{ Multi-Signature} \\
  \midrule
    Security & Low & Medium & High & High \\
    Usability & High & Medium & Low & High \\
    Examples & Metamask~\cite{metamask} & Ledger~\cite{ledger2018} &  & Safe~\cite{safe} \\
\end{tabular}
}
\label{tab:Wallet Security}
\end{center}
\end{table}
\vspace{-0.3in}

Blockchain wallets have four different types as below (Table~\ref{tab:Wallet Security}). 

\begin{itemize}

\item 
Softerware wallets: Softerware wallets are apps that users can use on the computer, phone, or Web browser. They are convenient because they let users manage their funds from anywhere as long as an internet connection exists. For example, mobile wallets include Trust Wallet~\cite{trustwallet} and MetaMask~\cite{metamask}, while desktop wallets include Exodus~\cite{exodus} and Atomic Wallet~\cite{atomicwallet}.

\item
Hardware wallets: Hardware wallets are physical devices that securely store the private keys offline in the encrypted chip, which provides an extra layer of security against hacking and malicious attacks. Users can connect the hardware wallet to a computer or mobile device by cable or Bluetooth when they needed to authorize transactions. For example, some popular hardware wallets include Ledger Nano S Plus~\cite{ledger}, Trezor~\cite{trezor}, and KeepKey~\cite{keepkey}.

\item
Paper wallets: A paper wallet is a physical document that prints or writes the public address and private key on paper. Paper wallets are considered one of the most secure forms of storage since they are immune to online hacking. However, they require careful protection from physical damage.

\item
Multi-sig wallet: It is one type of software wallets. The wallet requires multiple private keys to authorize transactions, adding an extra layer of security and reducing the risk of unauthorized access~\cite{han2021efficient}. These wallets are suitable for organizations that control funds among multiple parties.

\end{itemize}

For creating a new wallet, there are some wallet standards (i.e., Bitcoin Improvement Proposal, BIP), which define technical specifications for wallet implementation. Common standards include BIP-39 (Mnemonic code for generating deterministic keys)~\cite{Marek2013}, BIP-32 (Hierarchical Deterministic Wallets)~\cite{Pieter2012}, and BIP-44 (Multi-Account Hierarchy for Deterministic Wallets)~\cite{Marek2014}.

Blockchain wallets play a crucial role in facilitating the adoption, and use of cryptocurrencies and digital assets. They empower individuals to take control of their financial assets, enabling peer-to-peer transactions, investment in DeFi and NFT platforms, and participation in blockchain-based ecosystems.

\subsection{Mainstream Blockchain Platforms}

As already mentioned at the start of this study, Bitcoin, Ethereum, Cosmos, Polkadot, Optimism,  Arbitrum and Solana are mainstream blockchain platforms. Herewith we provide detailed explanations and summarise them with different properties~\cite{agbo2019comparison}, as shown in Table~\ref{table:Blockchains Property}.  

Bitcoin was introduced in 2009~\cite{bitcoin0block} by an anonymous individual or group under the pseudonym Satoshi Nakamoto. It is the world's first and most well-known cryptocurrency~\cite{nakamoto2008peer}. It operates on a decentralized peer-to-peer network, allowing users to transact directly without the need for intermediaries like banks or payment processors. Bitcoin's primary use case is a digital currency or store of value, with transactions recorded on a public ledger. One of Bitcoin's key features is its decentralized nature, maintained by a distributed network of miners who validate and secure transactions through a process called mining. Bitcoin uses a PoW consensus mechanism, where miners compete to solve complex mathematical puzzles to add new blocks to the blockchain and receive rewards in the form of newly minted bitcoins. On the other hand, Bitcoin has a a maximum supply capped at 21 million coins, which causes Bitcoin's scarcity. Moreover, this Bitcoin's scarcity is also from its decentralized architecture and fixed supply schedule. Due to Bitcoin’s scarcity, it has been viewed as a hedge against inflation and store of value similar to gold~\cite{kyriazis2020bitcoin}.

Ethereum was launched in 2015 by Vitalik Buterin and a team of developers. It is a decentralized blockchain platform that enables the creation of smart contracts and dApps~\cite{wood2014ethereum}. Ethereum serves as a programmable blockchain platform that allows developers to build and deploy dApps, which differentiates it from Bitcoin that focuses on digital currency. One of Ethereum's most significant innovations is the introduction of smart contracts, self-executing contracts with the terms of the agreement directly written into code. Smart contracts enable a wide range of use cases, including DeFi, NFT, tokenization, digital identity, gaming, and more. Ethereum uses a different consensus mechanism compared to Bitcoin, transitioning from PoW to PoS with the upcoming Ethereum 2.0 upgrade. PoS aims to improve scalability, security, and energy efficiency by allowing validators to create new blocks and secure the network based on the amount of tokens they hold and are willing to "stake" as collateral~\cite{sanka2021systematic}.

Cosmos is a decentralized network of independent blockchains, each powered by Byzantine Fault Tolerant (BFT) consensus algorithms like Tendermint. The Cosmos Network aims to create an "Internet of Blockchains," where diverse blockchains can connect and exchange assets and data seamlessly. This approach enhances scalability by allowing multiple blockchains to run in parallel while maintaining sovereignty and security~\cite{scoville2007cosmic}.

Polkadot is a multi-chain network that enables interoperability between different blockchains. This architecture enhances scalability by allowing transactions to be processed in parallel across multiple chains~\cite{burdges2020overview}. Polkadot also introduces the concept of "bridges" to connect external blockchains, enabling seamless cross-chain communication and asset transfers.

Optimism processes multiple transactions independently on Optimism, instead of Ethereum. After the transactions are finalized, a single "optimistic" proof is submitted to the Ethereum mainnet. This mechanism processes transactions significantly faster and with cheaper gas fees compared to executing them directly on the Ethereum network. Optimism blockchain can balance between scalability and security, leveraging Ethereum's security guarantees while reducing its congestion issues~\cite{li2023security}.

\vspace{-0.2in}
\begin{table}[!htbp]
\caption{Mainsteam Blockchains Platforms}
\begin{center}
\vspace{-0.2in}
\begin{tabular}{c|c c c c c c c}
   \textbf{ Blockchains } & \textbf{ Security} & \textbf{ Consensus} &\textbf{Speed} & \textbf{ Throughput} & \textbf{ Transaction Cost} \\ 
   \midrule
    Bitcoin & High & PoW & Slow & Low & High \\
    Ethereum & Medium & PoW and PoS & Medium & Medium & Medium& \\
    Cosmos & Depends & BFT  & Fast & High & Low & \\
    Polkadot & Depends & NPoS & Fast & High & Low & \\
    Optimism & Depends & PoH and PoS & Fast & High & Low & \\
    Arbitrum & Depends & PoW and PoS & Fast & High & Low & \\
    Solana & Depends & PoW and PoS & Fast & High & Low & \\
\end{tabular}
\label{table:Blockchains Property}
\end{center}
\end{table}
\vspace{-0.4in}

Arbitrum blockchain works similar to Optimism, Arbitrum employs Optimistic Rollups protocol, called Arbitrum Rollup, to process transactions in Arbitrum blockchain and then submit the proof to the Ethereum mainnet. However, Arbitrum's AVM is designed to be compatible with Ethereum's existing smart contracts, which makes it easier for dApp developers to migrate their applications to Arbitrum blockchain. Through the seamless experiences for smart contract developers, Arbitrum aims to improve the scalability of the Ethereum ecosystem and accelerate the adoption of layer-2 scaling solutions.

Solana distinguishes itself from others by offering fast transaction speeds, low fees, and high throughput. Their properties are making it an attractive option for developers seeking to build scalable and efficient dApps. Solana is using a different consensus mechanism called PoH. PoH is a cryptographic clock that timestamps each transaction before it enters the network, ensuring a secure and immutable record of transactions while enabling fast transaction processing. By combining PoH with PoS consensus mechanism, Solana can achieve high throughput and low latency and process thousands of transactions per second. Solana's architecture focus on horizontal scaling through its multi-threaded transaction processing approach. This allows Solana to handle a large number of transactions simultaneously~\cite{pierro2022can}, making it well-suited for applications that require high-speed and high-volume transaction processing, such as decentralized exchanges, gaming platforms, and financial applications.

Examples of all the above mainstream blockchain platforms can be found in the evolution of NFTs, which is discussed below. 

\section{NFT-Non Fungible Token}
\subsection{NFT History}

The first NFT, Quantum~\cite{quantum}, was minted in 2014 on Namecoin~\cite{kalodner2015empirical} by Kevin McCoy, but the full potential of NFTs was realized with the Ethereum blockchain, which offered a more reliable and accessible platform for launching NFT projects. Over time, NFTs have been used to represent a variety of assets, including real estate, digital art, and gaming tokens. In 2015, EverdreamSoft created “Spells of Genesis,”~\cite{spellsofgenesis} the first blockchain trading card game, on top of Bitcoin, pioneering true ownership of digital assets in gaming. The same year saw the creation of Etheria~\cite{etheria}, a so-called blocky realm where players could become real estate tycoons. In 2017, Larva Labs launched CryptoPunks~\cite{cryptopunks}, one of the earliest NFT generative art collections, which influenced popular projects like Bored Ape Yacht Club. Dapper Labs’ CryptoKitties~\cite{cryptokitties}, also from 2017, was one of the first blockchain games on Ethereum and inspired the ERC-721 standard. Axie Infinity~\cite{axieinfinity}, created by Sky Mavis in 2018, became a popular blockchain game where players collect and trade digital creatures called Axies. Decentraland~\cite{decentraland}, launched in 2020, allows users to buy and sell virtual land and items. NBA Top Shot~\cite{nbatopshot} transformed sports footage into digital collectibles, while Art Blocks~\cite{artblocks} , also from 2020, streamlined the creation of generative art NFTs. Bored Ape Yacht Club, founded in 2021 by Yuga Labs, became a critically acclaimed PFP NFT collection. Azuki~\cite{azuki}, which emerged in early 2022, quickly became a front runner in the PFP-centric NFT market. In 2024, Pandora~\cite{pandora} introduced the ERC-404 standard, where tokens can be regenerated with new characteristics when moved between wallets, blending the uniqueness of NFT technology with the liquidity traits of fungible tokens.

\vspace{-0.2in}
\begin{figure}[!htp]
    \centering
    \includegraphics[width=0.95\linewidth]{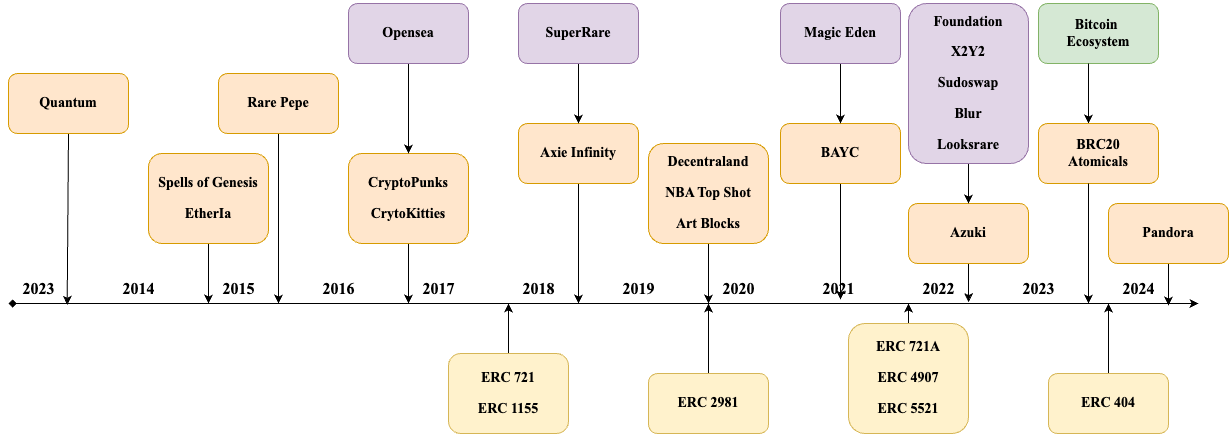}
    \caption{NFT History}
    \label{fig:nft history}
\end{figure}
\vspace{-0.2in}

An upsurge in NFT activities expanded beyond the Ethereum ecosystem to include the Bitcoin ecosystem~\cite{kiraz2023nft} through a technique known as \textit{inscription}~\cite{li2024bitcoin}. Due to significant structural and architectural differences, Bitcoin-based NFTs require additional development efforts~\cite{wang2023understanding,wang2024bridging}, particularly because of limitations inherent in the UTXO model. Additionally, trading NFTs on the Bitcoin network often incurs higher transaction (i.e. gas) fees. The Figure \ref{fig:nft history} shows the timeline for NFT history, NFT standards and NFT markets.

Next, we explore the concept of NFTs, highlighting their technical affordances and examining their connections to visual arts in the pre-blockchain era.

\subsection{NFT Overview}
\noindent\textbf{Deployers and Users.}
In the overall landscape of NFTs, two key roles emerge: the deployer and the user. The deployer, often an artist, creator, or technologist, wields the ability to manifest digital assets into the blockchain realm. As the central deployer in this process, they orchestrate the creation and deployment of NFT smart contracts, sculpting the essence of token standards, metadata intricacies, and additional functionalities. They usually have the admin permission for the NFT Smart Contracts~\cite{wang2019blockchain}, which means they control the particularities of NFT Smart Contracts. With each deployment, the artwork is endowed a new digital ownership. The other key role is maintained by users, including collectors, investors, or enthusiasts. As expected, they browse NFT marketplaces and platforms to discover digital art, collectibles, and virtual assets, interacting with NFT smart contracts with normal permissions. Through their activities, they give these NFT value and create new stories about ownership and cultural importance, which are often shared or even celebrated on other social media platforms such as LinkedIn, TikTok, Facebook, and X (formerly Twitter).

\begin{figure}[!htp]
    \centering
    \includegraphics[width=0.8\linewidth]{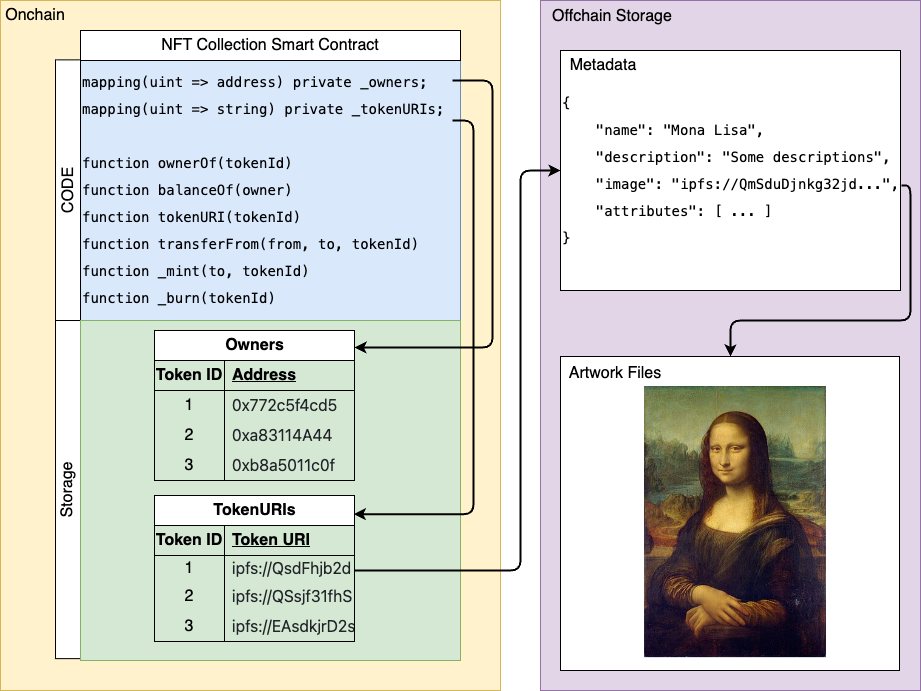}
    \caption{NFT Structure}
    \label{fig:nft structure}
\end{figure}

\smallskip
\noindent\textbf{NFT Structure.}
NFTs have a comprehensive structure that defines how they work (cf. Figure~\ref{fig:nft structure}). Each NFT belongs to a collection and has a unique identifier in the collection called a token ID, which differentiates it from others. Every token ID has its own metadata, which includes important properties like the name, description, creator, image, and other relevant information about the "minted" digital asset This metadata helps explain the NFT’s background in detail.

NFT related transactions on the blockchain record and track the information concerning who owns each NFT, as well as each of its subsequent owners. This ensures transparency and immutability. Smart contracts provide different functions of the NFT, such as creation, transfer, royalty charging, and other features that benefit both creators and collectors.

Furthermore, NFTs often use off-chain storage to store metadata and digital assets. There is a link between NFT and metadata called \emph{tokenURI}, which is a standard way to link to the metadata and digital asset information. The \emph{tokenURI} usually includes a URL that points to where the metadata are stored, which makes it easy for users and apps to access and show NFT's information. Some developers save the encoded metadata with Base64 into \emph{tokenURI}. Users need to view information after decoding with Base64.

\noindent\textbf{Metadata.}
NFT metadata is at the core of defining the characteristics of NFTs~\cite{barrington2022role}. NFTs represent unique digital assets on the blockchain network that are verified and tradable, with their token ID and information primarily determined by metadata. The token ID ensures that each token is unique and identifiable, while the metadata describes the attributes and characteristics of the NFT. Although metadata is usually stored off-chain, it is linked to the NFT via the \emph{tokenURI} on the blockchain. Metadata includes various attributes such as the title, description, image URI, creator information, and other attributes. The title provides the name of the NFT and artwork, while the description explains what the NFT represents in terms of the artwork. The image URI links to the digital art or digital assets stored in other storage. Metadata is typically formatted in JSON~\cite{pezoa2016foundations}, which is easy to read and parse. This structured format is crucial not only for human readability but also for machine compatibility. For storage services, centralized storage servers can be used, but decentralized storage solutions like InterPlanetary File System (IPFS)~\cite{benet2014ipfs} are preferred due to their resilience and resistance to censorship.

\subsection{NFT Lifecycle}
\begin{figure}[htp]
    \centering
    \includegraphics[width=0.95\linewidth]{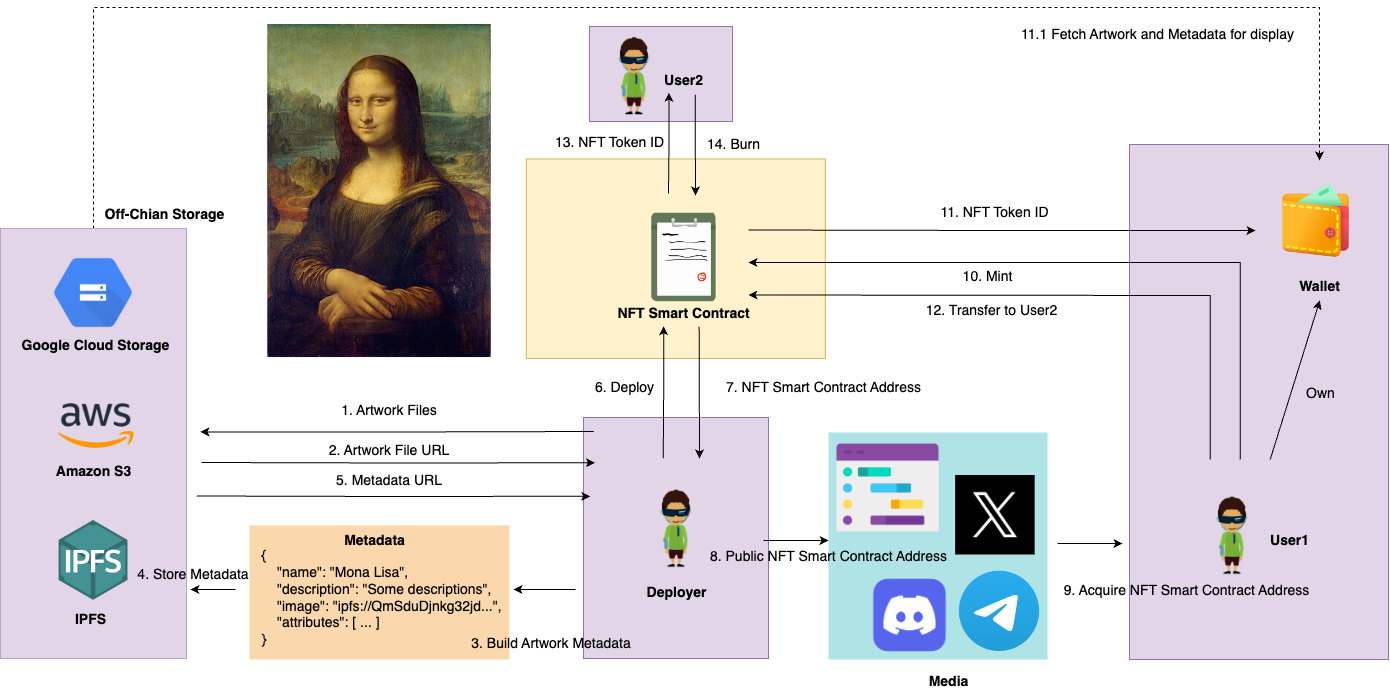}
    \caption{NFT Lifecycle}
    \label{fig:nft lifecycle}
    \vspace{-0.2in}
\end{figure}

The lifecycle of an NFT involves several stages~\cite{yang2023non} (cf. Figure~\ref{fig:nft lifecycle}).

\smallskip
\noindent\textbf{Creation.}
During this phase, the digital asset is tokenized on a blockchain network, using smart contracts to ensure its uniqueness and ownership status.

Five steps to create or tokenize an NFT:

\begin{itemize}[itemsep=0pt,topsep=0pt,parsep=0pt]

\item 
Deployer stores the artwork files in off-chain storage, such as AWS S3~\cite{amazons3}, Google Cloud~\cite{cloudgoogle} or IPFS, and acquires the artwork's URL.

\item 
Deployer builds the metadata for the NFT collections, stores the metadata files in off-chain storage and acquires the metadata's URL. 

\item 
Deployer deploys the smart contract for the NFT collection. One NFT Smart contract actually is one NFT collection and there are many NFTs inside. Each NFT has its own unique token ID. And for the NFT collection, the unique ID is the smart contract address. Deployer needs to set metadata URL for each NFT according NFT token ID, which is devised to build relationship between metadata and NFT. 

\item 
Deployer publicises this NFT Smart Contract address to the public through the project's website and social media, such as Twitter (now rebranded as X)~\cite{x}, Telegram~\cite{telegram}, Discord ~\cite{discord}, etc.

\item 
Users can \emph{mint} their own NFT after they acquire an NFT Smart Contract address. They can \emph{mint} one or many NFTs depending on the limitation settings in the NFT Smart Contract established by the deployer. There are two primary ways to mint a digital asset. First, users can mint by themselves. Second, the deployer mints and then airdrops an NFT to one or more users. 

Specific details regarding the actual minting process are discussed shortly. Suffice it to say that ownership, storage and display, transfer, and destruction are key elements in the NFT ecosystem. 

\begin{minted}
{solidity}
function mint(address to) external;
\end{minted}
\end{itemize}

\smallskip
\noindent\textbf{Ownership.}
The buyer owns NFT and can display it in their digital wallets or showcase it in virtual environments, depending on the type of asset. Ownership is verified through cryptographic keys associated with the buyer's wallet address.

\begin{itemize}
\item[] 
\begin{minted}[breaklines]
{solidity}
function ownerOf(uint256 _tokenId) external view returns (address);
\end{minted}
\end{itemize}

\smallskip
\noindent\textbf{Storage and Display.}
NFTs can be stored in digital wallets that support the specific blockchain network on which they were created. Additionally, many NFTs are displayed in virtual galleries, online platforms, or even in metaverse environments~\cite{khati2022non}, where they can be showcased to a wider audience.

\smallskip
\noindent\textbf{Transfer.}
NFTs can be transferred to new owners through subsequent sales or gifted to others. As they are built on the blockchain technology, records of ownership are maintained indefinitely, allowing for a clear lineage of ownership over time. Owners can transfer their NFT via \emph{transferFrom}  method with from, to, and \emph{tokenId} parameters in the NFT smart contract.

\begin{itemize}
\item[] 
\begin{minted}[breaklines]
{solidity}
function transferFrom(address _from, address _to, uint256 _tokenId) external;
\end{minted}
\end{itemize}

\smallskip
\noindent\textbf{Destruction.}
In some cases, creators may choose to retire their NFTs, removing them from circulation to increase the rarity of existing tokens. This can happen for various reasons, such as the end of a limited edition series or changes in the creator's preferences. Deployer or users can destroy their NFT by the \emph{burn} method with their NFT \emph{tokenId}. However, some NFT is build with permission control, the \emph{burn} method might be limited to use for users.

Next, we discuss NFT standards, which offer good interoperability, standardized interfaces, higher security, increased liquidity, and management efficiency.

\begin{itemize}
\item[] 
\begin{minted}{solidity}
function burn(uint256 tokenId) external;
\end{minted}
\end{itemize}

\subsection{NFT Standards}

Table~\ref{tab:NFT Standards Property} summarizes the properties of the NFT standards.

\vspace{-0.1in}
\begin{table*}[!htbp]
\caption{NFT Standards Property}
\vspace{-0.3in}
\begin{center}
\resizebox{1\linewidth}{!}{
\begin{tabular}{c|c  c  c  c  c  c  c  c}

\toprule
  \textbf{ Feature }  
 & \rotatebox{0}{\textbf{ ERC-721}} 
 & \rotatebox{0}{\textbf{ 721A}} 
 & \rotatebox{0}{\textbf{ 1155}}
 & \rotatebox{0}{\textbf{ 2981}}
 & \rotatebox{0}{\textbf{ 4907}} 
 & \rotatebox{0}{\textbf{ 3525}} 
 & \rotatebox{0}{\textbf{ 5521}} 
 & \rotatebox{0}{\textbf{ 404 }} \\
 \midrule

Batch Transfer?	& \xmark & \cmark & \cmark & \xmark & \xmark & \xmark & \xmark & \xmark \\
 
Easily Query Owner?  & \cmark& \cmark & \xmark & \cmark & \cmark & \cmark & \cmark & \cmark \\
 
 Support semi-fungible? & \xmark & \xmark & \cmark & \xmark & \xmark & \cmark & \xmark & \cmark \\ 
 
 Compatible (ERC20)? & \xmark & \xmark & \cmark & \xmark & \xmark & \cmark & \xmark & \cmark \\  \midrule

  Gas Efficiency & Low & High & Depends & Low & Low & Low & Low & Low \\ \midrule
  
 Use cases & \makecell{Artwork, \\CryptoKitties~\cite{cryptokitties}, \\BAYC~\cite{boredapeyachtclub}} & \makecell{Collectibles, \\Azuki~\cite{azuki}} & \makecell{Enjin~\cite{enjin}, \\Gaming, \\Artwork} & & & Solv~\cite{solv} & & Pandora~\cite{pandora} \\

 \bottomrule
\end{tabular}
}
\label{tab:NFT Standards Property}
\end{center}
\end{table*}
\vspace{-0.3in}

\noindent\textbf{ERC-721: Non-Fungible Token Standard.}
ERC-721~\cite{William2018} is a fundamental standard in the blockchain space, specifically designed for creating and managing NFTs on the Ethereum network. It is introduced in 2018 by William Entriken and colleagues. ERC-721 establishes rules for generating unique tokens that represent unique assets, unlike fungible tokens such as cryptocurrencies. This standard enables functionalities such as ownership tracking, secure transfer of NFTs, and approval mechanisms for third-party management, etc. Additionally, ERC-721 connects metadata with each token, enhancing their uniqueness and value. By providing a standard framework, ERC-721 has been involved in the widespread adoption and interoperability of NFTs, helping the growth of digital art and other unique digital assets within the blockchain ecosystem.

\smallskip
\noindent\textbf{ERC-1155: Multi Token Standard.}
ERC-1155~\cite{Witek2018} is an innovative standard in the blockchain world that allows for the creation and management of both fungible tokens and NFTs within a single smart contract on the Ethereum blockchain network. It was introduced by Witek Radomski and his team in 2018 to improve some features based on earlier standards like ERC-20 and ERC-721. It allows developers to manage ERC-20 and ERC-721 tokens more efficiently within a single contract. This means they can create complex systems for things like in-game items, digital art, and other digital assets. ERC-1155 also lets you send multiple tokens in one transaction, which cuts down on transaction costs. 

\smallskip
\noindent\textbf{ERC-721A.}
ERC-721A~\cite{James2022} is a proposed extension of the ERC-721 standard. The purpose of ERC-721A is to enhance the functionality and usability of NFTs on the Ethereum blockchain. ERC-721A is proposed by James Sangalli and the Ethereum community. Moreover, ERC-721A has additional functions to ERC-721, such as batch minting, which executes the creation of multiple tokens in a single transaction. The purpose is to improve efficiency and reducing gas costs. 

\smallskip
\noindent\textbf{ERC-2981: NFT Royalty Standard.}
ERC-2981~\cite{Zach2018} is a proposed Ethereum Improvement Proposal (EIP), also known as "EIP-2981", which introduces a standardized interface for royalties on NFTs. ERC-2981 is proposed by William Entriken and David Moore in 2021. ERC-2981 helps to compensates creators and original rights holders for their artwork in the NFT ecosystem. This standard imports a new feature that allows for the transparent distribution of royalties to the original creators whenever NFTs are resold on the secondary markets.

\smallskip
\noindent\textbf{ERC-3525: Semi-Fungible Token.}
ERC-3525~\cite{Wang2020} bridges the gap between fungible (replaceable) and non-fungible (unique) tokens by allowing tokens to have both types of properties, making them partially fungible. The main features of ERC-3525 are its composability, interoperability, and enhanced flexibility. When applied to visual art, ERC-3525 can change how digital art is created, owned, and traded. It supports partial ownership, meaning artwork can be tokenized and divided into smaller parts, so multiple investors can own a share. This helps artists manage licensing and royalties more effectively, ensuring they get fair profits when their work is used or sold. Additionally, ERC-3525 increases the liquidity of visual art, making it easier to trade and invest in high-value pieces.

\smallskip
\noindent\textbf{ERC-4907: Rental NFT, an Extension of EIP-721.}
ERC-4907~\cite{Anders2018} is another extension of the popular ERC-721 standard that introduces rental features for NFTs. It adds a new role called ‘user,’ which can be assigned to an address and has a set expiration time. The ‘user’ role allows someone to use the NFT but does not give them the ability to transfer it or change the permissions. This feature makes it easy to rent out NFTs by default, so artists don’t have to manually manage rental periods or worry about their artwork being returned safely.

\smallskip
\noindent\textbf{ERC-5521: Referable NFT.}
ERC-5521~\cite{Saber2022} is an innovative standard in the Ethereum network that extends the popular ERC-721 standard. ERC-5521 introduces two referable indicators, known as '\emph{referring}' and '\emph{referred}', and a time-based indicator called '\emph{createdTimestamp}'. These new features allow the formation of a directed acyclic graph (DAG) relationship between each NFTs. This type of relationship benefits in that it enables artists to query, track, and analyze their relationships within the network easily.

\smallskip
\noindent\textbf{ERC-404.}
ERC-404~\cite{Keir2024} is an experimental token standard on the Ethereum blockchain. This standard is introducing a new class of assets, combining the characteristics of both ERC-20 (fungible tokens) and ERC-721 (NFTs). The core of ERC-404 lies in its innovative approach to bring fungibility to NFTs. ERC-404 uses a mint and burn strategy to manage the non-fungibility of NFTs. In addition, ERC-404 creates a working combination between ERC-721 and ERC-20 that are proportionally tied to each other. This token standard allows for NFTs to be fractionalized within the boundaries of the Ethereum protocol. This fractional ownership is embedded directly into the NFT’s smart contract which provides function for the conversion between NFTs and tokens.

Then, we focuse on the actual process in making an NFT.

\section{Applying NFTs to Visual Art}

\subsection{Procedures for NFT Minting and Trading in the Marketplace}
NFTs are unique digital assets that are traded on various online marketplaces~\cite{white2022characterizing,jo2021efficient}. Here's a general overview of how NFT trading typically works in Figure~\ref{fig:nft trading}:

\begin{itemize}
\setlength{\itemsep}{0pt}

\item 
\textbf{Creation.} An artist or creator produces a digital asset, such as artwork, music, videos, virtual real estate, or other digital collectibles. They then mint this asset into an NFT on a compatible blockchain platform. This process essentially turns the digital file into a one-of-a-kind token, making it unique and verifiable. Creator need to list the NFT Smart Contract in their project website or NFT marketplace to show it to public users.

\item 
\textbf{Mint.} User can mint their NFT from the project website or NFT marketplace. In gernal, NFT marketplace have friendly tools to help creators deploy and mint a new NFT.

\begin{figure}[!htp]
    \centering
    \includegraphics[width=\linewidth]{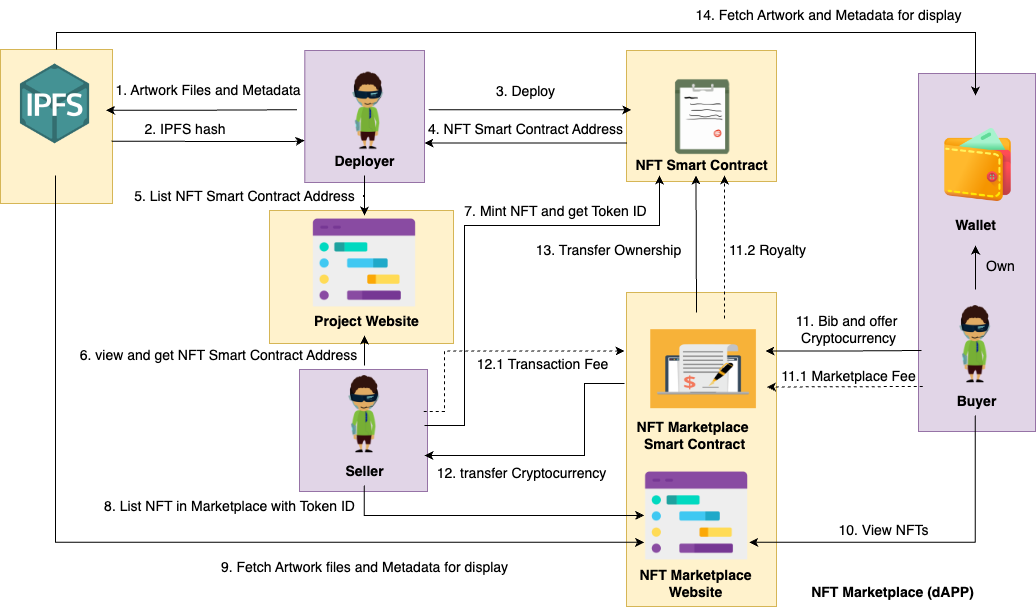}
    \caption{NFT Trading}
    \label{fig:nft trading}
\end{figure}

\item 
\textbf{Listing on Marketplaces.} Once minted, the NFT is listed for sale on a marketplace, which can be one of the decentralized platforms like OpenSea, Rarible, or Foundation, or they can be centralized platforms like Nifty Gateway or NBA Top Shot. Each platform may have its own requirements and procedures for listing NFTs.

\item \textbf{Ownership Transfer.} When someone purchases an NFT, the ownership of that digital asset is transferred to their digital wallet. The transaction is recorded on the blockchain, providing an immutable record of ownership.

\item 
\textbf{Payment.} Buyers typically pay for NFTs using cryptocurrency, usually via Ethereum (ETH) on Ethereum-based NFT platforms. Some platforms may also support other cryptocurrencies. The cryptocurrency will be paid to seller through NFT marketplace smart contract. The transaction fee or marketplace fee is charged directly by the charge logic designed by creator in the NFT marketplace smart contract.

\item 
\textbf{NFT Smart Contracts.} NFT contracts, which are self-executing contracts with the terms of the agreement directly written into code, govern the transfer of NFT ownership and the distribution of proceeds from sales. These contracts ensure that the creator receives royalties for subsequent resales of the NFT, typically a percentage of the sale price.

\item 
\textbf{Market Dynamics.} NFT prices can fluctuate based on demand, perceived value, and the reputation of the creator. Factors such as rarity, uniqueness, and the creator's reputation can all influence the price of an NFT~\cite{mekacher2022rarity}.

\item 
\textbf{Storage and Display.} After purchasing an NFT, owners can store them in compatible digital wallets. Some platforms also offer ways to display NFTs in virtual galleries~\cite{ivancic2016virtual} or digital environments.

\end{itemize}

It is important to research and understand the specific marketplace and blockchain platform before participating in NFT trading to ensure security, authenticity, and compliance with any applicable regulations.

\subsection{NFT Marketplace and Price}
\subsubsection{Marketplaces}

The Cryptopunks Marketplace~\cite{punksforsale} is the prime of the earlier platform for buying, selling, and trading Cryptopunks, one and most iconic NFT projects. Created by Larva Labs, Cryptopunks are 10,000 unique digital collectible characters, each with distinct traits and features such as hairstyles, accessories, and backgrounds. The marketplace operates as a decentralized exchange where users can browse through the available Cryptopunks and purchase them using cryptocurrency, typically Ethereum. Sellers can list their Cryptopunks for sale, setting their own prices, and buyers can acquire these digital collectibles either for investment purposes or for their personal collections. Due to their scarcity and historical significance in the NFT space, Cryptopunks have garnered significant attention and value within the digital art and crypto communities. The marketplace provides a platform for collectors to acquire these rare digital assets and for sellers to monetize their holdings. 

OpenSea~\cite{opensea} is one of the largest and most prominent decentralized marketplaces for buying, selling, and trading a wide variety of digital assets, primarily focusing on NFTs. Launched in 2017, OpenSea has become a gateway to the NFT world and community. It offers a wide range of digital collections, including artwork, virtual real estate, domains, and more. OpenSea allows users to create, buy, and sell NFTs through a decentralized platform empowered by blockchain technology. It is supported on multiple blockchains, such as Ethereum and Polygon~\cite{polygon}, providing greater flexibility and accessibility. The platform’s main feature is its user-friendly interface, which simplifies the process of browsing, buying, and managing NFTs. Users can explore new collections, discover digital assets, and interact with other collectors. OpenSea also provides tools for users to mint their own NFTs, enabling artists, game developers, and content creators to tokenize their creations and sell them directly to a global audience. Additionally, OpenSea offers features such as auctions, bundles, and peer-to-peer trading, enhancing the trading experience for users.

\vspace{-0.2in}
\begin{figure}[!htp]
    \centering
    \includegraphics[width=\linewidth]{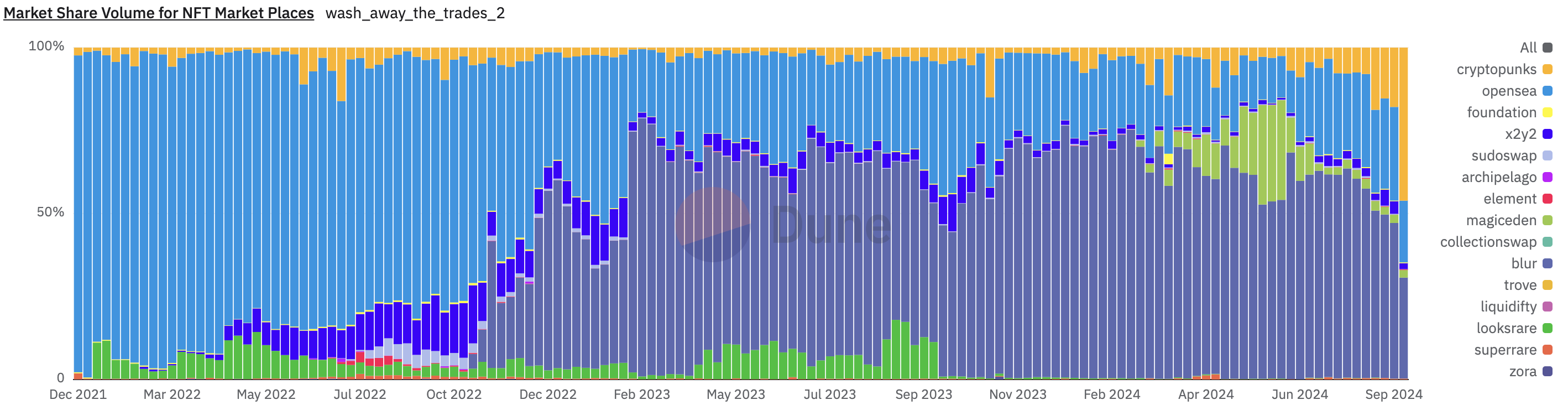}
    \caption{NFT Marketplace Share~\cite{nftmarketplacecomparison}}
    \label{fig:NFT Marketplace Comparison}
\end{figure}
\vspace{-0.2in}

Foundation~\cite{foundation} is a curated marketplace and community-driven platform for buying, selling, and minting digital art NFTs. Launched in February 2021, it quickly gained attention for its focus on quality, creativity, and empowering artists within the NFT space. Unlike some other NFT platforms, Foundation operates on an invite-only basis, indicating artists need to be nominated or invited by existing members to join. This curation process ensures that only high-quality and innovative artwork is showcased, helping Foundation build its reputation as a top destination for digital art collectors and enthusiasts. This platform features a user-friendly interface that allows users to browse and discover digital art in various styles, mediums, and themes. Each artwork listed on Foundation is represented as an NFT and backed by blockchain technology to guarantee authenticity, provenance, and ownership. One of Foundation’s unique features is its auction-based model, where artists can choose to auction their digital art to the highest bidder. This dynamic auction system enhances engagement, excitement, and fair market value for creators and collectors alike. Additionally, Foundation provides tools and resources for artists to mint their own NFTs directly on the platform, enabling them to tokenize and monetize their digital creations while maintaining control over their intellectual property.

X2Y2~\cite{x2y2} is an innovative platform focused on creating, exploring, and trading NFTs. It aims to democratize access to NFTs and provides a space for creators and collectors to interact with various digital assets. On X2Y2, artists and creators can mint their own NFTs and showcase their work to a global audience. This includes everything from generative art to various types of digital collectibles and tokenized assets. One of X2Y2’s key features is its emphasis on community and collaboration. The platform fosters a supportive environment where artists can share ideas, work on projects together, and learn from each other. This collaborative spirit has helped build a diverse and vibrant NFT community within the X2Y2 ecosystem. X2Y2 also offers tools and resources that make it easy for artists to create and mint their NFTs directly on the platform. This streamlined process allows artists to tokenize their work and sell it to collectors, opening up new opportunities for monetization and exposure. For collectors, X2Y2 provides a curated marketplace where they can discover and acquire NFTs from talented artists around the world.

Sudoswap~\cite{sudoswap} is a DEX built on blockchain technology, specifically designed for users within the Ethereum ecosystem. It’s known for enabling trustless swaps of NFTs on the Ethereum blockchain. Sudoswap supports both ERC721 and ERC1155 NFTs, as well as all ETH and ERC20 tokens. As a decentralized exchange, Sudoswap operates without a central authority, allowing users to control their own funds and conduct peer-to-peer transactions through smart contracts. Compared to traditional centralized exchanges, this decentralized approach offers several advantages, including enhanced security, lower fees, and greater transparency. One of Sudoswap’s key features is its minimalist, low gas-efficient automated market maker (AMM) protocol~\cite{xu2023sok}. This protocol uses customizable bonding curves to facilitate the swapping of NFTs and tokens, and vice versa. In addition to its core trading functions, Sudoswap also offers liquidity pools~\cite{gupta2023role} where users can contribute their tokens and earn rewards through liquidity mining. This mechanism helps incentivize liquidity on the platform, leading to deeper liquidity and improved trading efficiency.

Blur~\cite{blur2024} is an innovative NFT marketplace that caters to professional traders. Launched in 2022, Blur quickly surpassed OpenSea in terms of volume due to its low fees, high speed, and unique features such as NFT floor depth charts. Blur is more than just a marketplace; it's an ecosystem consisting of an NFT marketplace, aggregator~\cite{cousaert2022sok}, and lending platform~\cite{darshan2023blockchain}. The Blur marketplace provides advanced analytics, portfolio management features, and the ability to compare NFTs on different marketplaces. The marketplace's main feature is that it allows for gasless transactions, which significantly reduces transaction costs for users~\cite{prechtel2019evaluating}. To get started, users first connect their wallet and sign a gasless transaction. Then, they start listing and bidding on NFTs. The platform encourages users to upload or list NFTs to qualify for airdrops. Blur has its native token, which has been distributed to the community over several quarters.

\begin{table*}[htbp]
\caption{NFT Marketplace Features}
\resizebox{\linewidth}{!}{
\begin{tabular}{c|c|c|c|c|c|c|c|c|c}
\toprule

\textbf{ Features}              & \textbf{ Cryptopunks } & \textbf{ Opensea } & \textbf{ Foundation} &\textbf{  X2Y2 }   & \textbf{ Sudoswap }    & \textbf{ Blur }  & \textbf{ Superrare} & \textbf{ Looksrare} & \textbf{  MagicEden } \\

\midrule

\makecell{Royalty}     & 0\%         & 0.5\%   & 10\%       & 0\%     & 0\%          & 0.5\%                        & 10\%      & 0\%       & 0\%        \\

\makecell{MarketFee}      & 0\%         & 0\%     & 5\%        & 0\%     & 0\%          & 0\%                          & 3\%       & 2\%       & 2\%        \\

\midrule

\makecell{Support \\Chains} &
  Ethereum &
  \makecell{Ethereum, \\Polygon, \\Arbitrum, \\Avalanche, \\Base, Blast, \\Optimism, \\Klaytn} &
  Ethereum &
  Ethereum &
  \makecell{Ethereum, \\Base, \\Arbitrum, \\Blast} &
  \makecell{Ethereum, \\Blast} &
  Ethereum &
  Ethereum &
  \makecell{Solana, \\Bitcoin, \\Ethereum, \\Polygon, \\Base} \\ 
  \midrule
Mint                 & \xmark          & \cmark     & \cmark        & \cmark     & \xmark           & \cmark                          & \cmark       & \cmark       & \cmark        \\
Auction              & \xmark          & \cmark     & \cmark        & \xmark      & \cmark          & \xmark                           & \cmark       & \xmark        & \cmark        \\
Drops                & \xmark          & \cmark     & \cmark        & \xmark      & \xmark           & \cmark                          & \xmark        & \xmark        & \cmark        \\
Bundles              & \xmark          & \cmark     & \xmark         & \xmark      & \cmark          & \xmark                           & \xmark        & \xmark        & \xmark         \\
\makecell{P2P\\Trading} & \xmark          & \cmark     & \xmark         & \xmark      & \cmark          & \xmark                           & \xmark        & \xmark        & \xmark         \\
\midrule

\makecell{Support \\Standards} &
  ERC-721 &
  \makecell{ERC-721, \\ERC-1155} &
  \makecell{ERC-721, \\ERC-1155} &
  \makecell{ERC-721, \\ERC-1155} &
  \makecell{ERC-721, \\ERC-1155, \\ERC-20} &
  \makecell{ERC-721, \\ERC-1155} &
  \makecell{ERC-721, \\ERC-1155} &
  \makecell{ERC-721, \\ERC-1155, \\ERC-2981} &
  \makecell{ERC-721, \\ERC-1155} \\
  \midrule
\makecell{Utility Token}        &             &         &            & X2Y2    & SUDO         & BLUR                         & RARE      & LOOKS     &            \\
\midrule
Others               &             &         &            & Lending & AMM Protocol & \makecell{Aggregator, \\Lending, \\Gasless} &           &           & Launchpad \\
 
\bottomrule
\end{tabular}
}
\label{tab:NFT Marketplace Features}
\vspace{-0.3in}
\end{table*}

SuperRare~\cite{superrare} is a leading platform for buying, selling, and collecting digital art as NFTs. Launched in 2018, it focuses on high-quality, limited-edition digital artwork created by both emerging and established artists. One of SuperRare’s key features is its curation process, which ensures that top-notch artwork is featured on the platform. Artists must apply or be invited to join SuperRare, and each piece of art goes through a thorough review by the platform’s curation team. This strict curation helps maintain the integrity of the art on SuperRare, making it a trusted marketplace for collectors. The platform offers a vibrant marketplace where users can discover and acquire digital art in various styles, media, and themes. SuperRare supports different formats, including static images, animated GIFs, and even interactive 3D experiences, allowing artists to explore new ways of expressing themselves digitally. A unique aspect of SuperRare is its focus on provenance and royalties. Every time an artwork is resold, a portion of the sale proceeds automatically goes to the original creator, ensuring that artists continue to benefit from the increasing value of their work over time.

LooksRare~\cite{looksrare} is a decentralized, community-driven NFT marketplace that was launched in January 2022 as an alternative to the established platforms like OpenSea. It quickly gained attention thanks to its unique approach and features tailored to the NFT community. One of the key aspects of LooksRare is its reward system. Buyers and sellers of NFTs earn LOOKS tokens, the platform’s native cryptocurrency, as rewards. These tokens can be staked for additional rewards, which encourages active participation. The platform also offers competitive trading fees, with a portion of the fees being redistributed to LOOKS stakers, further motivating community involvement. Additionally, LooksRare ensures that creators receive royalties on secondary sales, allowing them to continue earning from their work. The user experience is designed to be intuitive and seamless, making it easy for users to browse, buy, and sell NFTs. By focusing on rewarding active users, providing lower fees, and emphasizing decentralization, LooksRare has positioned itself as a compelling option in the growing NFT marketplace.

Magic Eden~\cite{magiceden} is a decentralized marketplace based on the Solana blockchain, focusing on buying, selling, and trading NFTs. Launched to meet the growing demand for NFTs and digital collectibles, Magic Eden offers a range of features designed to provide a smooth experience for users within the Solana ecosystem. At its core, Magic Eden is a marketplace where users can discover, buy, and sell various NFTs, including digital art, collectible cards, virtual items, and more. The platform has an intuitive interface that makes it easy to browse and explore different collections, helping users find unique and valuable NFTs that match their interests. One of Magic Eden’s main advantages is its integration with the Solana, known for its high performance, scalability, and low transaction fees. By leveraging Solana’s capabilities, Magic Eden provides fast and cost-effective transactions, allowing users to buy, sell, and trade NFTs with minimal friction. Additionally, Magic Eden offers features like staking, governance, launchpads, and community engagement. Users can stake SOL tokens to earn rewards, participate in governance decisions, and contribute to the platform’s growth.

Table~\ref{tab:NFT Marketplace Features} shows the main features for the NFT marketplace. The Figure \ref{fig:NFT Marketplace Comparison} provides the NFT marketplace share from Dec 2021 to Sep 2024.

\smallskip
\noindent\textbf{NFT Pricing.}
NFTs represent a unique class of digital assets, often associated with digital art, collectibles, and virtual real estate. Unlike fungible cryptocurrencies, each NFT is distinct, often carrying unique metadata that determines its individual value. Moreover, there is much research on NFT prices, as shown in Table \ref{tab:Summary of the Identified Literature for NFT Price}. Understanding the price determinants of NFTs involves different factors that includes textual and visual data, market dynamics, social media influence, and broader financial trends. Textual descriptions and image data within NFT collections can help explain price variations among individual NFTs; however, features extracted from these data do not generalize well to new, unseen collections~\cite{ziemke_what_2023}. This indicates that, while certain stylistic or descriptive elements might boost an NFT's value within a specific collection, these factors do not necessarily apply across different collections. Interestingly, NFTs described with negative sentiment tend to fetch higher prices than those with positive sentiment~\cite{othman_impact_2023}. This could be attributed to the perceived uniqueness or controversial nature of such descriptions, which might attract more attention and demand.

\begin{table*}[!htbp]
\centering
\caption{Summary of the Identified Literature for NFT Price}
\resizebox{\linewidth}{!}{
\begin{tabular}{p{1.3in}|p{1.2in}|p{1.2in}|p{1in}|p{0.3in}|p{1.2in}|p{1.2in}|p{1.2in}}
  \toprule
\multicolumn{1}{c}{\textbf{Title}} &
\textbf{Data Source} &
\textbf{Time Period} &
\textbf{Topic} &
\textbf{AI Used} &
\textbf{AI Model } &
\textbf{Analysis} &
\textbf{Data Analysis} \\
  \midrule  
What Determines the Price of NFTs?~\cite{ziemke_what_2023} &
  OpenSea &
  January 2018 to May 2023 &
  Influences NFT pricing &
  \cmark &
  LR, RR, LASSO, DT. VGG16, ResNet50, etc. &
  Machine Learning and CNN &
  The 140 Collection by Twitter, Google Trends interest metrics. \\\hline
Impact of Crypto Art Sentiment on Art Valuation~\cite{othman_impact_2023} &
  OpenSea &
  January 2022 to March 2022 &
  Influences NFT pricing &
  \xmark &
   &
  VADER (Valence Aware Dictionary and Sentiment Reasoner) &
  NFT asset name, description, date of transaction, and transaction price. \\\hline
Non-Fungible Tokens: What Makes Them Valuable?~\cite{leitter_non-fungible_2023} &
  Twitter &
  1 June 2023 to 1 July 2023 &
  Influences NFT pricing &
  \cmark &
  Symbolic AI, Subsymbolic AI, Neurosymbolic AI &
  Natural Language Processing (NLP) &
  200,000 tweets \\\hline
Non-fungible token (NFT) markets on the Ethereum blockchain: temporal development, cointegration and interrelations~\cite{ante2023non} &
  Ethereum Blockchain &
  June 2017 and May 2021 &
  Influences NFT pricing &
  \xmark &
   &
  Vector Autoregression (VAR) model. &
  NFT sales, dollar volume of NFT trades, the number of unique blockchain wallets that traded NFTs \\\hline
Prediction and interpretation of daily NFT and DeFi prices dynamics: Inspection through ensemble machine learning \& XAI~\cite{ghosh_prediction_2023} &
  www.investing.com, coinmarketcap.com. &
  1 January 2020 to 31 July 2022 &
  Prodict NFT price &
  \cmark &
  Isometric Mapping and Uniform Manifold Approximation and Projection techniques, Gradient Boosting Regression and Random Forest &
  Machine learning, Explainable Artificial Intelligence (XAI) &
  DeFi tokens price and NFT tokens price \\\hline
The relationship between trading volume, volatility and returns of Non-Fungible Tokens: evidence from a quantile approach~\cite{yousaf_relationship_2022} &
  coinmarketcap.com &
  17 January 2018 to 20 November 2021 &
  Influences NFT pricing &
  \xmark &
   &
  Quantile VAR approach &
  Daily data of volume and prices of three NFTs, including THETA, Tezos (XTZ), Enjin Coin (ENJ) \\\hline
Dynamic dependence and predictability between volume and return of Non-Fungible Tokens (NFTs): The roles of market factors and geopolitical risks~\cite{urom_dynamic_2022} &
  nonfungible.com &
  23 June 2017 to 11 February 2022 &
  Influences NFT pricing &
  \xmark &
   &
  Quantile cross-spectral coherency and quantile regression techniques. &
  Series plots of prices and volumes for the NFT market, three NFT sub-markets including Decentraland, CryptoKitties and Cryptopunks. \\\hline
Volatility spillovers across NFTs news attention and financial markets~\cite{wang2022volatility} &
  LexisNexis News \& Business database, nonfungible.com, coinmarketcap.com, &
  5 January 2018 to 3 June 2022 &
  Prodict NFT price &
  \xmark &
   &
  TVP-VAR, Pooled OLS, GARCH-MIDAS &
  590m news stories \\\hline
Price determinants of non-fungible tokens in the digital art market~\cite{horky_price_2022} &
  SuperRare, Kaggle.com &
  March 2020 to March 2021 &
  Prodict NFT price &
  \cmark &
  Machine learning: Hedonic pricing models &
  Machine learning &
  Description, name, size, and type and covers \\\hline
Understanding digital bubbles amidst the COVID-19 pandemic: Evidence from DeFi and NFTs~\cite{maouchi_understanding_2022} &
  coinmarketcap.com, defipulse.com, policyuncertainty.com &
  22 January 2020 to 15 March 2021 &
  Influences NFT pricing &
  \xmark &
   &
  Real-time bubble detection method &
  9 DeFi tokens, 3 NFTs, Bitcoin, and Ethereum, \\\hline
The economic value of NFT: Evidence from a portfolio analysis using mean–variance framework~\cite{ko_economic_2022} &
  Yahoo Finance, CoinMarketCap &
  4 December 2019 to 9 June 2021 &
  Influences NFT pricing &
  \xmark &
   &
  mean–variance approach &
  The stock market index, bond index, US dollar index, commodity index, and the cryptocurrency price \\\hline
NFTs and asset class spillovers: Lessons from the period around the COVID-19 pandemic~\cite{aharon_nfts_2022} &
  investing.com, nonfungible.com &
  1 January 2018 to 30 June 2021 &
  Influences NFT pricing &
  \xmark &
   &
  Time-Varying Parameter Vector Autoregressions (TVPVAR) approach &
  Gold, equities, currencies, bonds, cryptocurrencies \\\hline
The NFT Hype: What Draws Attention to Non-Fungible Tokens?~\cite{pinto-gutierrez_nft_2022} &
  trends.google.com, nonfungible.com, coinmarketcap.com, finance.yahoo.com &
  1 December 2017 to 30 July 2021 &
  Influences NFT pricing &
  \xmark &
   &
  Vector autoregressive (VAR) models &
  Bitcoin and Ethereum prices, VIX index, S\&P 500 index, and gold prices \\\hline
Is non-fungible token pricing driven by cryptocurrencies?~\cite{dowling_is_2022} &
  coinmarketcap.com, nonfungible.com, &
  March 2019 to March 2021 &
  Influences NFT pricing &
  \xmark &
   &
  Volatility spillover methodology, cross-waveletsfollowing the approach &
  Bitcoin and Ether, Decentraland LAND tokens; CryptoPunk images; and Axie Infinity game. \\\hline
Fertile LAND: Pricing non-fungible tokens~\cite{dowling2022fertile} &
  nonfungible.com &
  March 2019 to March 2021 &
  Influences NFT pricing &
  \xmark &
   &
  Automatic variance ratio test, Automatic portmanteau test, Consistent test &
  4936 trades of LAND in our dataset \\
  \bottomrule
\end{tabular}
}
\label{tab:Summary of the Identified Literature for NFT Price}
\end{table*}

The trading volume of an NFT collection often correlates with its online presence. Metrics such as social media followers and website traffic can serve as indicators of an NFT's popularity and market activity~\cite{ziemke_what_2023}. High online engagement can boost visibility and drive demand, subsequently influencing prices. NFT markets exhibit significant interconnectedness with other financial markets, including cryptocurrencies, equity, and commodities. Volume and volatility spillovers between NFTs and these markets are particularly pronounced in extreme market conditions, with asymmetrical connectedness observed in upper and lower quantiles~\cite{yousaf_relationship_2022}. This indicates that extremely volatile market movements in NFTs are often influenced by broader financial market dynamics. The success of newer NFT projects is influenced by the performance of more established markets. Conversely, the growth and innovation in newer NFT markets can also impact the established projects~\cite{ante2023non}. This bidirectional influence highlights the interdependence within the NFT ecosystem.

NFT markets are relatively immature and often inefficient~\cite{dowling2022fertile}. For instance, the pricing of virtual real estate in Decentraland does not always reflect all available information, though a general rise in value has been observed over time. This inefficiency suggests opportunities for speculative gains but also underscores the risks involved. NFTs have been found to offer significant diversification benefits within traditional asset-based portfolios~\cite{ko_economic_2022}. During stable market times, NFTs can act as transmitters of systemic risk, but in the stressful periods, they tend to absorb risk spillovers~\cite{aharon_nfts_2022}. This dual role highlights their potential as both a hedge and a diversification tool in investment portfolios. The popularity of NFTs are heavily influenced by the performance of major cryptocurrencies like BTC and Ether~\cite{pinto-gutierrez_nft_2022}. Increased prices could drive hype and attention towards NFTs, as evidenced by co-movement patterns observed through wavelet coherence analysis~\cite{dowling_is_2022}. Furthermore, Bitcoin returns have been shown to predict NFT growth in popularity, as measured by Google search queries.

The key factors contributing to NFT value have been identified: rarity, reputation, authenticity, significance, utility, interoperability, demand, provenance, royalties, and hype~\cite{leitter_non-fungible_2023}. These aspects encapsulate both intrinsic qualities of NFTs and external market perceptions that drive their valuation. During extreme market conditions, various economic indicators such as equity and gold market uncertainties, business conditions, and geopolitical risks significantly predict returns in specific NFT markets~\cite{urom_dynamic_2022}. This indicates that broader economic factors can play a crucial role in determining NFT prices. The valuation of NFTs is influenced by a complex interplay of textual and visual data, social media presence, market dynamics, financial contagion, and broader economic conditions. These factors collectively shape the perceived value of NFTs.

\subsection{NFT Security}
\begin{figure}[htp]
    \centering
    \includegraphics[width=\linewidth]{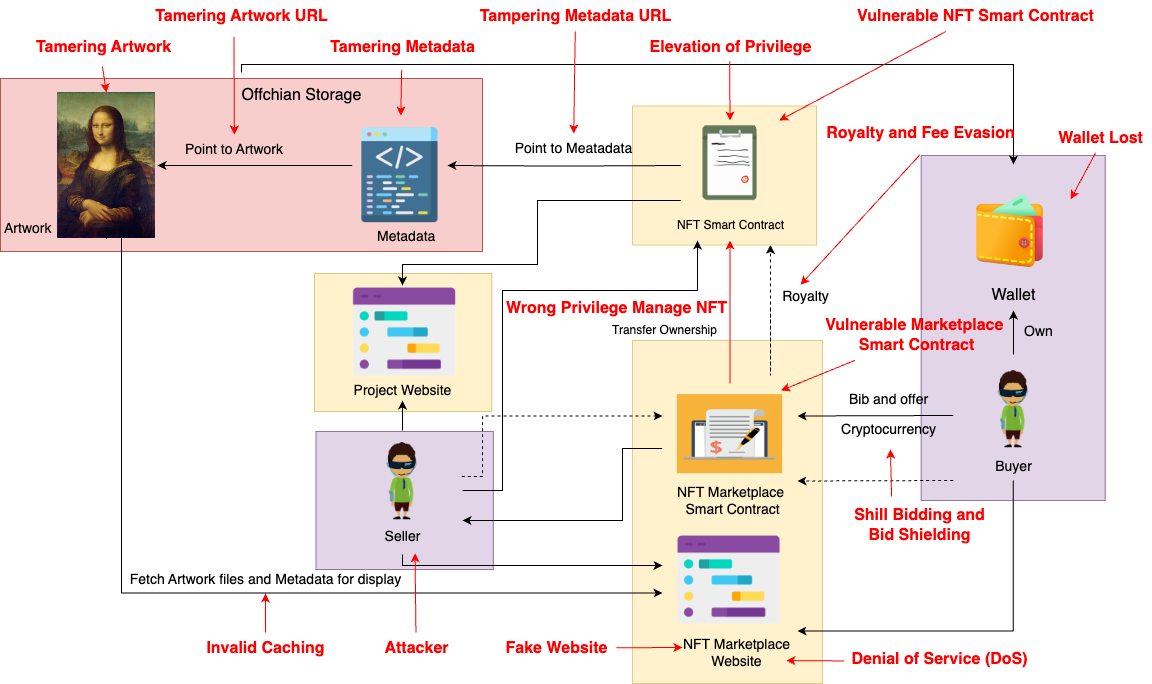}
    \caption{NFT Security Issues}
    \label{fig:nft security issues}
    \vspace{-0.3in}
\end{figure}

The rapid proliferation of NFTs market is facing with challenges in cybersecurity that pose risks to artists and collectors alike, including both human and technological issues. Figure~\ref{fig:nft security issues} shows most of the security issues for NFT trading.

On the human front, artists and collectors face threats from fraud, theft, and scams~\cite{roy2024unveiling}. Phishing attacks are particularly prevalent, where scammers deceive users into revealing sensitive information or performing actions that compromise their digital wallets. By impersonating legitimate services or using deceptive messages, scammers can trick users into confirming on-chain operations that transfer NFTs and other tokens directly to the attackers. This can result in substantial financial losses and the unauthorized transfer of valuable digital assets.

Technological vulnerabilities are common because of the complex nature of smart contracts and the intricate design of NFTs. Smart contracts, which are self-executing contracts with the terms of the agreement directly written into code~\cite{khan2021blockchain}, are foundation to the operation of NFTs. However, they are not immune to security flaws. Identity verification issues might allow malicious attackers to impersonate legitimate users and lead to unauthorized transactions. The verifiability of NFT contracts is another critical issue. NFT contract is authentic and has not been tampered with is essential for maintaining trust in the system. Smart contract vulnerabilities, such as reentrancy, transaction order dependence, integer overflows, and unhandled exceptions, all highlight the security challenges between NFTs and broader DeFi ecosystems~\cite{bose2022sailfish,frank2020ethbmc,grech2018madmax,jiang2018contractfuzzer,kalra2018zeus,luu2016making}.

Tampering with NFT metadata presents another significant risk. Metadata can be altered if proper security measures are not in place. This can lead to losing connection between the digital asset and its provenance. Invalid caching, where outdated or incorrect data is stored and used, can further complicate the matters by providing false information about the NFT~\cite{das2022understanding}.

In addition, NFT market suffers from a lack of transparency~\cite{cornelius2021betraying}, particularly in off-chain trades where transactions occur outside the blockchain. This opacity can facilitate fraudulent activities and make it difficult to verify the authenticity and ownership history of an NFT. Fairness in bidding processes is another concern, with practices such as shill bidding and bid shielding undermining the integrity of auctions. Royalty and marketplace fee evasion pose financial risks to artists and platforms, while cross-platform issues and non-enforcement of contractual terms can further exacerbate financial risks.

Security issues in Web2, such as Denial of Service (DoS) attacks~\cite{needham1993denial}, also persist in the NFT ecosystem. With substantial funds flowing into DeFi applications, the allure of scams and cyber attacks has grown. Cryptoeconomic strategies, such as transaction reordering~\cite{kelkar2020order,li2023transaction}, flash loan abuse~\cite{qin2021attacking}, arbitrage opportunities~\cite{daian2020flash}, and pump-and-dump schemes~\cite{xu2019anatomy,kamps2018moon}, have been adapted to exploit vulnerabilities in the NFT market. 

This intricate web of human and technological vulnerabilities underscores the urgent need for robust security measures to protect artists and their digital assets in the burgeoning NFT marketplace. Ensuring the verifiability of token contracts and implementing the principle of least privilege in smart contracts can address several technological vulnerabilities~\cite{xu2018blendcac}. Transparency in transactions, fairness in bidding, and robust enforcement of contractual terms are essential for maintaining trust and integrity in the NFT ecosystem~\cite{das2022understanding}. As the market continues to grow, ongoing efforts to identify and mitigate security risks will be crucial in safeguarding the interests of all participants in the NFT space.

\vspace{-0.1in}
\begin{figure}[!htp]
    \centering
    \includegraphics[width=\linewidth]{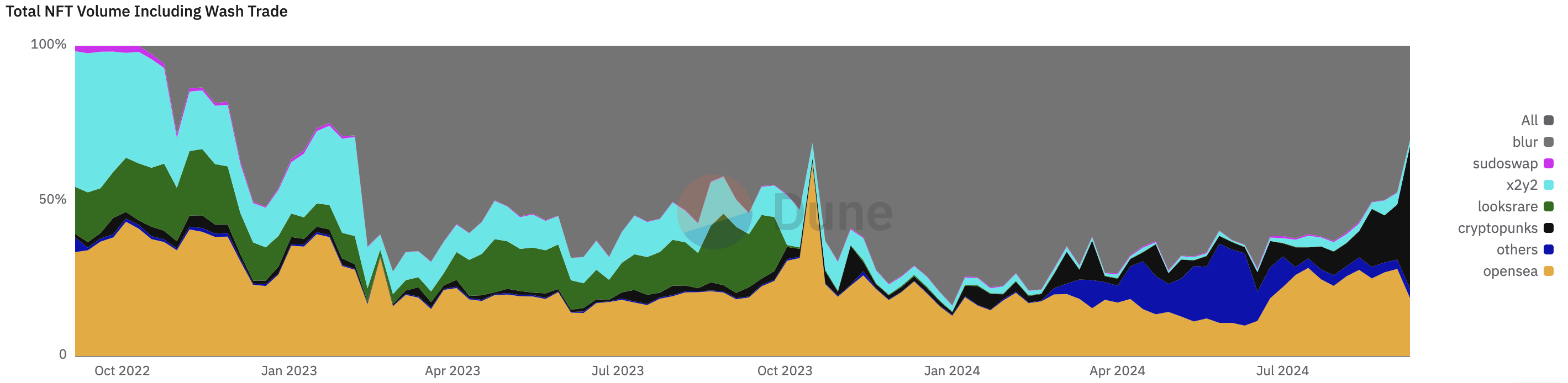}
    \caption{NFT Wash Trading~\cite{nftwashtrading}}
    \label{fig:Wash Trading}
\end{figure}

\noindent\textbf{Wash Trading.} The emergence of NFTs has ushered in a new era of digital asset markets, characterized by unprecedented innovation and rapid growth. However, amidst the excitement surrounding NFTs lies a darker underbelly: the prevalence of wash trading (cf. Figure~\ref{fig:Wash Trading}). Wash trading, a deceptive practice involving the simultaneous buying and selling of assets to create artificial trading volume, has raised significant concerns within the NFT ecosystem. Recent studies challenged previous estimates, suggesting that wash trading may be less common than initially thought~\cite{von_wachter_nft_2022}. Nonetheless, findings still indicate alarming levels of wash trading activity, with some estimates it attributing up to 25\% of total trading volume to this illicit practice~\cite{tosic_beyond_2023}.

The scale of the issue becomes particularly evident when examining specific NFT collections. Shockingly, certain collections, such as Meebits, have seen as much as 93\% of their total trade volume linked to wash trading~\cite{tosic_beyond_2023}. Furthermore, wash trading isn't confined to obscure corners of the market; dominant centralized players are implicated, contributing to suspicions of market abuse. Incentivized structures within NFT markets exacerbate the problem, with markets like LooksRare and X2Y2 exhibiting wash trading volume proportions as high as 94.5\% and 84.2\%, respectively~\cite{niu_unveiling_2024}.

Identifying and combating wash trading in NFT markets presents unique challenges. Innovative solutions are being proposed, such as algorithms to detect wash trading at different levels, from transaction to collector~\cite{tahmasbi_identifying_2024}. These efforts aim to provide regulators and investors with tools to navigate the complexities of the NFT landscape and mitigate the risks associated with market manipulation. Moreover, the motivations behind wash trading are multifaceted. While some engage in wash trading to artificially inflate NFT prices, others exploit token reward systems offered by marketplaces 
~\cite{morgia_game_2023}. The profitability of wash trading is evident, with some perpetrators avoiding significant losses and even profiting from reselling NFTs to unsuspecting buyers.As the NFT market continues to evolve, the need for robust regulation becomes increasingly apparent. Measures must be implemented to prevent market abuse, instill investor confidence, and safeguard the integrity of digital asset markets. Novel visualization tools, like NFTDisk~\cite{wen_nftdisk_2023}, offer investors a means to identify wash trading activities and make informed decisions in an increasingly complex market landscape. Table~\ref{tab:Summary of the identified literature for NFT Wash Trading} provides the summary of the identified literature for NFT Wash Trading.

\vspace{-0.2in}
\begin{table*}[!htbp]
\caption{Summary of the identified literature for NFT Wash Trading}
\resizebox{\linewidth}{!}{
\begin{tabular}{p{2.5in}|c|c|l|c} 

\toprule
\textbf{Title} & \textbf{NFT Marketplaces} &
  \textbf{Standards} &
  \textbf{Topic Category} &
  \textbf{Publications} \\
\midrule

NFT Wash Trading Quantifying suspicious behaviour in NFT markets~\cite{von_wachter_nft_2022} &
  OpenSea &
 \makecell{ERC-721 \\ ERC-1155} &
  Wash trading analysis &
  arXiv \\\midrule
Unveiling the Paradox of NFT Prosperity~\cite{huang_unveiling_2024} &
   \makecell{OpenSea,  CryptoPunks, \\ LooksRare, X2Y2, Blur }&
   \makecell{ERC-721 \\ ERC-1155} &
  Wash trading analysis &
  ACM \\\midrule
Unveiling NFT Wash Trading in Markets~\cite{niu_unveiling_2024} &
  X2Y2, LooksRare &
   &
  Wash trading analysis &
  ACM \\\midrule
Identifying Washtrading Cases in NFT Sales Networks~\cite{tahmasbi_identifying_2024} &
  Nifty Gateway &
   &
  Wash trading detection &
  IEEE Transactions \\\midrule
  
Suspicious trading in NFTs~\cite{sifat_suspicious_2024} &
  OpenSea &
   &
  Wash trading analysis &
  Information/Management \\ \midrule
  
Beyond the Surface: Advanced Wash Trading Detection in NFT Markets~\cite{tosic_beyond_2023} &
  Blur, OpenSea,  & \makecell{X2Y2 \\ ERC-721 \\ LooksRare} 
   &
  Wash trading analysis &
  arVix \\\midrule
The Dark Side of NFTs: A Large-Scale Empirical Study of Wash Trading~\cite{chen_dark_2023} &
  OpenSea &
  ERC-20 &
  Wash trading analysis &
  arVix \\\midrule
Crypto Wash Trading: Direct vs. Indirect Estimation~\cite{falk_crypto_2023} &
  LooksRare &
   &
  Wash trading analysis &
  arVix \\\midrule
A Game of NFTs: Characterizing NFT Wash Trading in the Ethereum Blockchain~\cite{morgia_game_2023} &
 \makecell{OpenSea, LooksRare,\\  Rarible, SuperRare, \\ Decentraland, Foundation} &
   \makecell{ERC-721 \\ERC-20} &
  Wash trading analysis &
  IEEE Conference \\\midrule
NFTDisk: Visual Detection of Wash Trading in NFT Markets~\cite{wen_nftdisk_2023} &
  OpenSea, LooksRare &
  ERC-721 &
  Wash trading detection &
  ACM   \\
  \bottomrule
\end{tabular}
}
\label{tab:Summary of the identified literature for NFT Wash Trading}
\vspace{-0.3in}
\end{table*}

\section{Discussion and Future Research}
\subsection{User Experience Improvement}
At the current stage, entering the NFT world presents significant challenges due to the complexity of websites and tools, such as digital wallets and related concepts, which differ fundamentally from traditional wallets and banking systems. These new concepts are entirely unfamiliar to the general public. Additionally, Web3 dApps require users to learn and practice new mechanisms~\cite{wang2023exploring} related to NFTs, creating barriers for both regular artists and users. Traditional notions and concepts do not translate well into the NFT world, necessitating the development of new tools with lower entry barriers and simplified concepts to enhance user experience within the NFT ecosystem.

To improve user experience, it is essential to streamline the onboarding process~\cite{yu2024toward}. This can begin with developing user-friendly interfaces that guide users through the setup of digital wallets and explain their function in an accessible manner. Tutorials and intuitive design are more clear to users and can help the process of creating and managing NFTs. Integrating familiar elements from traditional finance into wallets, such as simplified transaction overviews and clear, straightforward instructions, can help users execute the transition quickly and smoothly. Moreover, using ideas and tools from traditional systems in a way that fits with the NFT world can help bridge the gap.

The development of all-in-one platforms (i.e., integrating various NFT operations, such as creation and trading) can also enhance the user experience. These platforms should offer seamless interoperability with different blockchains and marketplaces by reducing the need for users to navigate multiple services. Some features can make NFTs more approachable, for example, automatic wallet setup, integrated educational resources, and user-friendly dashboards.

\subsection{Security and Privacy Improvement}
NFT benefits from the inherent transparency provided by blockchain technology. However, this transparency also raises potential issues. Smart contract audits by security teams can identify and prevent exploits. In addition, stronger authentication methods, such as multi-factor authentication (MFA), can significantly reduce the unauthorized access issues. For high value NFTs, cold storage, which means storing them in hardware wallets, can minimize hacking risks. Educating users about phishing scams and implementing anti-phishing measures are also crucial to protecting against attacks. Moreover, education should teach users to avoid giving away their private keys or other sensitive information. Implementing decentralized identity (DID)~\cite{reed2020decentralized} solutions can ensure secure and verifiable identities without relying on centralized entities, preventing identity fraud. Regular security updates for NFT platforms and tools are necessary to patch vulnerabilities and stay ahead of the latest security threats and best practices~\cite{choi2022future}.

Blockchain may compromise user privacy. For artists and NFT users, maintaining privacy is crucial~\cite{almashaqbeh2022sok}. Zero-knowledge proofs (ZKPs)~\cite{chaliasos2024sok} can verify transactions without revealing underlying details, ensuring that transactions are transparent and verifiable while keeping the identities of the parties involved private. Confidential transaction mechanisms~\cite{wang2020preserving} can hide transaction amounts and other sensitive details from the public while still ensuring validity. Creating private NFTs with privacy features, achieved through encryption and selective disclosure techniques, can ensure that only authorized parties can view metadata and ownership details. Off-chain privacy solutions, where sensitive information is stored off-chain (e.g., using trusted hardware~\cite{li2022sok}) while the blockchain is used only for validation, help maintain privacy by leveraging blockchain security while keeping private data off the public ledger. Developing and promoting the use of privacy-focused wallets with features like transaction mixing~\cite{tang2021analysis} and address obfuscation can help users maintain their anonymity.

The continuous involvement and collaboration of artists, users, governors, and developers are essential for the ecological perfection of the NFT ecosystem. Encouraging active community participation can lead to the early identification of issues and collaborative problem-solving. Bug bounty programs can incentivize security researchers to find and report vulnerabilities. Offering education and resources about best practices for security can help artists protect their assets. Moreover, rules are needed to protect users and encourage innovation from regulators which can help make the NFT space more secure.

\subsection{Remix: The Future of Next Generation NFTs}
Artwork remixing involves adding user's own ideas to an original artwork to create a new version. It is an interesting topic that benefits both artists and the NFT market. Artists can earn ongoing royalties whenever their work is resold as an NFT remix. It also lets users create new NFTs based on their own ideas, turning them into new artists. Theoretically, this remixing process can attract increasing numbers of active artists, as well as a variety of new creative outputs.

More specifically, artwork remixing~\cite{knobel2008remix} is promising as blockchain ensuring artists receive payment for every resale of their artwork. Original artists can keep earning money and get more recognition if more of their NFTs are traded. The continuous royalty payments incentivize artists to enter and engage in the NFT market.
For users, remixing and creating new versions of NFTs allow for personal expression and creativity. By turning existing artwork into new, unique pieces, users become co-creators and contribute to the diversity of the digital art world. This open approach to art creation means that anyone with interest and creativity can join the art community, potentially bringing in a lot of new talents and innovation to the art industry.

The marketplace also benefits because an increase in remixed artwork means an increased variety and volume of NFTs for trading. This variety can attract more buyers and collectors, making the market more lively and exciting. Original works can lead to many new derivative pieces, each with its own unique value. This ongoing process of creating and recreating helps build a strong and ever-evolving art ecosystem~\cite{waysdorf2021remix}.

\section{Conclusion}

NFT is still a new technology for stakeholders in the visual art sector and beyond. In this paper, we have surveyed the latest research on NFTs and their interrelated ecosystem. Firstly, we examined traditional artwork types, production, trading, and monetization methods. Secondly, we unpacked blockchain-related concepts such as structure, consensus algorithms, smart contracts, and digital wallets, and we compared and contrasted some well-known blockchain currencies. Thirdly, we reviewed the history of NFTs, their mechanisms, lifecycle, and standards on Ethereum. Fourthly, we investigates the potential for NFT artwork trading, including marketplaces, pricing, and security. Finally, we discussed some general findings and suggested areas for future research based on this aggregation and analysis of past scholarly, industry, and technical work.

\bibliographystyle{splncs04}
\bibliography{bib}

\begin{thebibliography}{100}
\providecommand{\url}[1]{\texttt{#1}}
\providecommand{\urlprefix}{URL }
\providecommand{\doi}[1]{https://doi.org/#1}

\bibitem{nftwashtrading}
21CO: {NFT} wash trading analysis. \url{https://dune.com/21co/wash-trading-in-nfts}  (2024)

\bibitem{aave}
Aave: Aave website. \url{https://aave.com}  (2024)

\bibitem{abidin2013printmaking}
Abidin, M.Z., Daud, W.S.A.W.M., Rathi, M.R.M.: Printmaking: Understanding the terminology. Procedia-Social and Behavioral Sciences  \textbf{90},  405--410 (2013)

\bibitem{adams2023uniswap}
Adams, H., Salem, M., Zinsmeister, N., Reynolds, S., Adams, A., Pote, W., Toda, M., Henshaw, A., Williams, E., Robinson, D.: Uniswap v4 core [draft] (2023)

\bibitem{adams2021uniswap}
Adams, H., Zinsmeister, N., Salem, M., Keefer, R., Robinson, D.: Uniswap v3 core. Tech. rep., Uniswap, Tech. Rep.  (2021)

\bibitem{etheria}
Adkisson, C.: Etheria world. \url{https://etheria.world}  (2024)

\bibitem{adler2018art}
Adler, A.: Why art does not need copyright. Geo. Wash. L. Rev.  \textbf{86}, ~313 (2018)

\bibitem{agbo2019comparison}
Agbo, C.C., Mahmoud, Q.H.: Comparison of blockchain frameworks for healthcare applications. Internet Technology Letters  \textbf{2}(5), ~e122 (2019)

\bibitem{aharon_nfts_2022}
Aharon, D.Y., Demir, E.: {NFTs} and asset class spillovers: Lessons from the period around the {COVID}-19 pandemic. Finance Research Letters  \textbf{47},  102515 (2022)

\bibitem{almashaqbeh2022sok}
Almashaqbeh, G., Solomon, R.: Sok: Privacy-preserving computing in the blockchain era. In: European Symposium on Security and Privacy (EuroSP). pp. 124--139. IEEE (2022)

\bibitem{amazons3}
Amazon: Amazon {S3} - cloud object storage. \url{https://aws.amazon.com/s3}  (2024)

\bibitem{Anders2018}
Anders, Lance, S.: {ERC}-4907: Rental {NFT}, an extension of {EIP}-721 (2018), ethereum Improvement Protocol, EIP-4907 \url{https://eips.ethereum.org/EIPS/eip-4907}

\bibitem{ante2023non}
Ante, L.: Non-fungible token ({NFT}) markets on the ethereum blockchain: Temporal development, cointegration and interrelations. Economics of Innovation and New Technology  \textbf{32}(8),  1216--1234 (2023)

\bibitem{saatchiart}
Art:, S.: Saatchi art. \url{https://www.saatchiart.com}  (2024)

\bibitem{artblocks}
ART~BLOCKS, I.: Art blocks website. \url{https://www.artblocks.io}  (2024)

\bibitem{azuki}
Azuki: Azuki website. \url{https://www.azuki.com}  (2024)

\bibitem{bahn1998cambridge}
Bahn, P.G.: The Cambridge illustrated history of prehistoric art. Cambridge University Press (1998)

\bibitem{barrington2022role}
Barrington, S.: The role of metadata in non-fungible tokens: Marketplace analysis and collection organization. arXiv preprint arXiv:2209.14395  (2022)

\bibitem{belchior2021survey}
Belchior, R., Vasconcelos, A., Guerreiro, S., Correia, M.: A survey on blockchain interoperability: Past, present, and future trends. ACM Computing Surveys (CSUR)  \textbf{54}(8),  1--41 (2021)

\bibitem{benet2014ipfs}
Benet, J.: Ipfs-content addressed, versioned, p2p file system. arXiv preprint arXiv:1407.3561  (2014)

\bibitem{bennett2013attractiveness}
Bennett, R., Kottasz, R.: Attractiveness of limited edition artwork for first-generation newly affluent consumers. International Journal of Arts Management  \textbf{15}(3), ~21 (2013)

\bibitem{bitshares}
Bitshares: Bitshares blockchain. \url{https://bitshares.github.io/}  (2024)

\bibitem{bitcoin0block}
blockchain: The genesis block of bitcoin. \url{https://www.blockchain.com/explorer/blocks/btc/0}  (2024)

\bibitem{blur2024}
Blur: Blur website. \url{https://blur.io}  (2024)

\bibitem{bonneau2015sok}
Bonneau, J., Miller, A., Clark, J., Narayanan, A., Kroll, J.A., Felten, E.W.: {SoK}: Research perspectives and challenges for {B}itcoin and cryptocurrencies. In: 2015 IEEE symposium on security and privacy. pp. 104--121. IEEE (2015)

\bibitem{bose2022sailfish}
Bose, P., Das, D., Chen, Y., Feng, Y., Kruegel, C., Vigna, G.: Sailfish: Vetting smart contract state-inconsistency bugs in seconds. In: IEEE Symposium on Security and Privacy (SP). pp. 161--178. IEEE (2022)

\bibitem{breton1969manifestoes}
Breton, A.: Manifestoes of surrealism, vol.~182. University of Michigan Press (1969)

\bibitem{brodie2014auction}
Brodie, N.: Auction houses and the antiquities trade. In: 3rd International Conference of Experts on The Return of Cultural Property. pp. 71--82. Archaeological Receipts Fund Athens (2014)

\bibitem{burdges2020overview}
Burdges, J., Cevallos, A., Czaban, P., Habermeier, R., Hosseini, S., Lama, F., Alper, H.K., Luo, X., Shirazi, F., Stewart, A., et~al.: Overview of polkadot and its design considerations. arXiv preprint arXiv:2005.13456  (2020)

\bibitem{canva}
Canva: Canva website. \url{https://www.canva.com}  (2024)

\bibitem{castro1999practical}
Castro, M., Liskov, B., et~al.: Practical byzantine fault tolerance. In: USENIX Symposium on Operating Systems Design and Implementation (OSDI). vol.~99, pp. 173--186 (1999)

\bibitem{chaliasos2024sok}
Chaliasos, S., Ernstberger, J., Theodore, D., Wong, D., Jahanara, M., Livshits, B.: {SoK}: What don't we know? understanding security vulnerabilities in {SNARK}s. USENIX Security Symposium (USENIX Security)  (2024)

\bibitem{chatzigiannis2022sok}
Chatzigiannis, P., Baldimtsi, F., Chalkias, K.: Sok: Blockchain light clients. In: International Conference on Financial Cryptography and Data Security (FC). pp. 615--641. Springer (2022)

\bibitem{chen_dark_2023}
Chen, S., Chen, J., Yu, J., Luo, X., Wang, Y., Zheng, Z.: The dark side of {NFTs}: A large-scale empirical study of wash trading, \url{http://arxiv.org/abs/2312.12544}

\bibitem{choi2022future}
Choi, M., Azzaoui, A., Singh, S.K., Salim, M.M., Jeremiah, S.R., Park, J.H.: The future of metaverse: Security issues, requirements, and solutions. Human-Centric Computing and Information Sciences  \textbf{12}(60),  1--14 (2022)

\bibitem{christies}
Christies: Christies website. \url{https://www.christies.com}  (2024)

\bibitem{quorum}
Consensys: Quorum project. \url{https://github.com/Consensys/quorum}  (2024)

\bibitem{cornelius2021betraying}
Cornelius, K.: Betraying blockchain: accountability, transparency and document standards for non-fungible tokens (nfts). Information  \textbf{12}(9), ~358 (2021)

\bibitem{cousaert2022sok}
Cousaert, S., Xu, J., Matsui, T.: {Sok}: Yield aggregators in defi. In: IEEE International Conference on Blockchain and Cryptocurrency (ICBC). pp. 1--14. IEEE (2022)

\bibitem{cryptokitties}
Cryptokitties: Cryptokitties website. \url{https://www.cryptokitties.co}  (2024)

\bibitem{punksforsale}
Cryptopunks: Cryptopunks: Punks for sale. \url{https://cryptopunks.app/cryptopunks/forsale}  (2024)

\bibitem{cryptopunks}
Cryptopunks: Cryptopunks website. \url{https://www.larvalabs.com/cryptopunks}  (2024)

\bibitem{curtis1997computer}
Curtis, C.J., Anderson, S.E., Seims, J.E., Fleischer, K.W., Salesin, D.H.: Computer-generated watercolor. In: Proceedings of the Annual Conference on Computer Graphics and Interactive Techniques (SIGGRAPH). pp. 421--430 (1997)

\bibitem{daian2020flash}
Daian, P., Goldfeder, S., Kell, T., Li, Y., Zhao, X., Bentov, I., Breidenbach, L., Juels, A.: Flash boys 2.0: Frontrunning in decentralized exchanges, miner extractable value, and consensus instability. In: IEEE symposium on security and privacy (SP). pp. 910--927. IEEE (2020)

\bibitem{dalla2021crowdfunding}
Dalla~Chiesa, C., Dekker, E.: Crowdfunding artists: beyond match-making on platforms. Socio-Economic Review  \textbf{19}(4),  1265--1290 (2021)

\bibitem{nbatopshot}
Dapper~Labs, I.: Nba top shot. \url{https://nbatopshot.com}  (2024)

\bibitem{darshan2023blockchain}
Darshan, M., Raswanth, S., Kumar, P., Srivastava, G.: A blockchain based framework for lending digital assets implemented using nft. In: IEEE IAS Global Conference on Emerging Technologies (GlobConET). pp.~1--6. IEEE (2023)

\bibitem{das2022understanding}
Das, D., Bose, P., Ruaro, N., Kruegel, C., Vigna, G.: Understanding security issues in the nft ecosystem. In: Proceedings of the 2022 ACM SIGSAC Conference on Computer and Communications Security. pp. 667--681 (2022)

\bibitem{de2018pbft}
De~Angelis, S., Aniello, L., Baldoni, R., Lombardi, F., Margheri, A., Sassone, V., et~al.: {PBFT} vs proof-of-authority: Applying the {CAP} theorem to permissioned blockchain. In: CEUR workshop proceedings. vol.~2058. CEUR-WS (2018)

\bibitem{decentraland}
Decentraland: Decentraland website. \url{https://decentraland.org}  (2024)

\bibitem{discord}
Discord: Discord website. \url{https://discord.com}  (2024)

\bibitem{dowling2022fertile}
Dowling, M.: Fertile {LAND}: Pricing non-fungible tokens. Finance Research Letters  \textbf{44},  102096 (2022)

\bibitem{dowling_is_2022}
Dowling, M.: Is non-fungible token pricing driven by cryptocurrencies? Finance Research Letters  \textbf{44},  102097 (2022)

\bibitem{dworkin2015sha}
Dworkin, M.J.: {SHA-3} standard: Permutation-based hash and extendable-output functions. National Institute of Standards and Technology  (2015)

\bibitem{elkins2011photography}
Elkins, J.: What photography is. Routledge (2011)

\bibitem{elkins2019painting}
Elkins, J.: What painting is. Routledge (2019)

\bibitem{enjin}
Enjin: Enjin website. \url{https://enjin.io/technology/wallet}  (2024)

\bibitem{ethash}
Ethereum: Ethash. \url{https://ethereum.org/en/developers/docs/consensus-mechanisms/pow/mining/mining-algorithms/ethash/}  (2024)

\bibitem{etsy}
Etsy: Etsy website. \url{https://www.etsy.com}  (2024)

\bibitem{spellsofgenesis}
EverdreamSoft: Spells of genesis. \url{https://spellsofgenesis.com}  (2024)

\bibitem{exodus}
Exodus: Exodus wallet. \url{https://www.exodus.com}  (2024)

\bibitem{facebook}
Facebook: Facebook website. \url{https://www.facebook.com}  (2024)

\bibitem{falk_crypto_2023}
Falk, B.H., Tsoukalas, G., Zhang, N.: Crypto wash trading: Direct vs. indirect estimation, \url{http://arxiv.org/abs/2311.18717}

\bibitem{far2022review}
Far, S.B., Bamakan, S.M.H., Qu, Q., Jiang, Q.: A review of non-fungible tokens applications in the real-world and metaverse. Procedia Computer Science  \textbf{214},  755--762 (2022)

\bibitem{ferguson1982defining}
Ferguson, S.C.: Defining the short story: impressionism and form. Modern Fiction Studies  \textbf{28}(1),  13--24 (1982)

\bibitem{Keir2024}
Finlow-Bates, K.: {ERC}-404 (2024), \url{https://github.com/Pandora-Labs-Org/erc404}

\bibitem{foundation}
Foundation: Foundation website. \url{https://foundation.app}  (2024)

\bibitem{frank2020ethbmc}
Frank, J., Aschermann, C., Holz, T.: $\{$ETHBMC$\}$: A bounded model checker for smart contracts. In: USENIX Security Symposium (USENIX Security). pp. 2757--2774 (2020)

\bibitem{fu2023ftx}
Fu, S., et~al.: Ftx collapse: a ponzi story. In: International Conference on Financial Cryptography and Data Security (FC). pp. 208--215. Springer (2023)

\bibitem{fu2024leveraging}
Fu, S., et~al.: Leveraging ponzi-like designs in stablecoins. International Journal of Network Management p. e2277 (2024)

\bibitem{gantefuhrer2004cubism}
Gantef{\"u}hrer-Trier, A.: Cubism. Taschen (2004)

\bibitem{ghelani2022non}
Ghelani, D.: What is non-fungible token ({NFT})? a short discussion about nft terms used in nft. Authorea Preprints  (2022)

\bibitem{ghosh_prediction_2023}
Ghosh, I., Alfaro-Cort{\'e}s, E., G{\'a}mez, M., Garc{\'\i}a-Rubio, N.: Prediction and interpretation of daily {NFT} and {DeFi} prices dynamics: Inspection through ensemble machine learning \& {XAI}. International Review of Financial Analysis  \textbf{87},  102558 (2023)

\bibitem{gilbert2003security}
Gilbert, H., Handschuh, H.: Security analysis of {SHA-256} and sisters. In: International workshop on selected areas in cryptography. pp. 175--193. Springer (2003)

\bibitem{cloudgoogle}
Google, G.: Cloud computing services | google google. \url{https://cloud.google.com}  (2024)

\bibitem{grech2018madmax}
Grech, N., Kong, M., Jurisevic, A., Brent, L., Scholz, B., Smaragdakis, Y.: Madmax: Surviving out-of-gas conditions in ethereum smart contracts. Proceedings of the ACM on Programming Languages (OOPSLA)  \textbf{2},  1--27 (2018)

\bibitem{groce2020actual}
Groce, A., Feist, J., Grieco, G., Colburn, M.: What are the actual flaws in important smart contracts (and how can we find them)? In: International Conference on Financial Cryptography and Data Security (FC). pp. 634--653. Springer (2020)

\bibitem{gudgeon2020sok}
Gudgeon, L., Moreno-Sanchez, P., Roos, S., McCorry, P., Gervais, A.: {SoK}: Layer-two blockchain protocols. In: International Conference on Financial Cryptography and Data Security (FC). pp. 201--226. Springer (2020)

\bibitem{gupta2023role}
Gupta, R., Gupta, M., Gupta, D.: Role of liquidity pool in stabilizing value of token. Scientific Journal of Metaverse and Blockchain Technologies  \textbf{1}(1),  9--17 (2023)

\bibitem{han2021efficient}
Han, J., Song, M., Eom, H., Son, Y.: An efficient multi-signature wallet in blockchain using bloom filter. In: Proceedings of the Annual ACM Symposium on Applied Computing (SAC). pp. 273--281 (2021)

\bibitem{hick2019artistic}
Hick, D.H.: Artistic license: the philosophical problems of copyright and appropriation. University of Chicago Press (2019)

\bibitem{horky_price_2022}
Horky, F., Rachel, C., Fidrmuc, J.: Price determinants of non-fungible tokens in the digital art market. Finance Research Letters  \textbf{48},  103007 (2022)

\bibitem{huang_unveiling_2024}
Huang, J., Xia, P., Li, J., Ma, K., Tyson, G., Luo, X., Wu, L., Zhou, Y., Cai, W., Wang, H.: Unveiling the paradox of nft prosperity. In: Proceedings of the ACM on Web Conference (WWW). pp. 167--177 (2024)

\bibitem{huang2019smart}
Huang, Y., Bian, Y., Li, R., Zhao, J.L., Shi, P.: Smart contract security: A software lifecycle perspective. IEEE Access  \textbf{7},  150184--150202 (2019)

\bibitem{hyperledgerfabric}
Hyperledger: Hyperledger fabric project. \url{https://bitshares.github.io/}  (2024)

\bibitem{axieinfinity}
Infinity, A.: Axie infinity whitepaper. \url{https://whitepaper.axieinfinity.com}  (2024)

\bibitem{instagram}
Instagram: Instagram website. \url{https://www.instagram.com}  (2024)

\bibitem{io2017eos}
IO, E.: {EOS}. {IO} technical white paper. EOS. IO (accessed 18 December 2017) https://github. com/EOSIO/Documentation  \textbf{6}, ~7 (2017)

\bibitem{ivancic2016virtual}
Ivancic, D., Schofield, D., Dethridge, L.: A virtual perspective: measuring engagement and perspective in virtual art galleries. International Journal of Arts and Technology  \textbf{9}(3),  273--298 (2016)

\bibitem{James2022}
James, S., Ethereum, c.: {ERC}-721{A} (2022), \url{https://github.com/chiru-labs/ERC721A}

\bibitem{nftmarketplacecomparison}
Jhackworth: Nft marketplace comparison. \url{https://dune.com/jhackworth/nftmarketplace}  (2024)

\bibitem{jiang2018contractfuzzer}
Jiang, B., Liu, Y., Chan, W.K.: Contractfuzzer: Fuzzing smart contracts for vulnerability detection. In: Proceedings of the ACM/IEEE International Conference on Automated Software Engineering (ASE). pp. 259--269 (2018)

\bibitem{jiang2023decentralized}
Jiang, E., Qin, B., et~al.: Decentralized finance ({DeFi}): A survey. arXiv preprint arXiv:2308.05282  (2023)

\bibitem{jo2021efficient}
Jo, K., Ko, J.: Efficient plan for art transaction through non fongible token ({NFT}). Asia-pacific Journal of Convergent Research Interchange  (2021)

\bibitem{kalodner2018arbitrum}
Kalodner, H., Goldfeder, S., Chen, X., Weinberg, S.M., Felten, E.W.: Arbitrum: Scalable, private smart contracts. In: USENIX Security Symposium (USENIX Security). pp. 1353--1370 (2018)

\bibitem{kalodner2015empirical}
Kalodner, H.A., Carlsten, M., Ellenbogen, P.M., Bonneau, J., Narayanan, A.: An empirical study of namecoin and lessons for decentralized namespace design. In: WEIS. vol.~1, pp. 1--23 (2015)

\bibitem{kalra2018zeus}
Kalra, S., Goel, S., Dhawan, M., Sharma, S.: Zeus: analyzing safety of smart contracts. In: Ndss. pp. 1--12 (2018)

\bibitem{kamps2018moon}
Kamps, J., Kleinberg, B.: To the moon: defining and detecting cryptocurrency pump-and-dumps. Crime Science  \textbf{7}(1),  1--18 (2018)

\bibitem{keepkey}
KeepKey: Keepkey website. \url{https://www.keepkey.com}  (2024)

\bibitem{kelkar2020order}
Kelkar, M., Zhang, F., Goldfeder, S., Juels, A.: Order-fairness for byzantine consensus. In: Annual International Cryptology Conference (CRYPTO). pp. 451--480. Springer (2020)

\bibitem{khan2021blockchain}
Khan, S.N., Loukil, F., Ghedira-Guegan, C., Benkhelifa, E., Bani-Hani, A.: Blockchain smart contracts: Applications, challenges, and future trends. Peer-to-peer Networking and Applications  \textbf{14},  2901--2925 (2021)

\bibitem{khati2022non}
Khati, P., Shrestha, A.K., Vassileva, J.: Non-fungible tokens applications: A systematic mapping review of academic research. In: 2022 IEEE 13th Annual Information Technology, Electronics and Mobile Communication Conference (IEMCON). pp. 0323--0330. IEEE (2022)

\bibitem{kickstarter}
Kickstarter: Kickstarter website. \url{https://www.kickstarter.com}  (2024)

\bibitem{kiraz2023nft}
Kiraz, M.S., Larraia, E., Vaughan, O.: {NFT} trades in {B}itcoin with off-chain receipts. International Conference on Applied Cryptography and Network Security (ACNS)  (2023)

\bibitem{klavans2014subject}
Klavans, J.L., LaPlante, R., Golbeck, J.: Subject matter categorization of tags applied to digital images from art museums. Journal of the Association for Information Science and Technology  \textbf{65}(1),  3--12 (2014)

\bibitem{knobel2008remix}
Knobel, M., Lankshear, C.: Remix: The art and craft of endless hybridization. Journal of adolescent \& adult literacy  \textbf{52}(1),  22--33 (2008)

\bibitem{ko_economic_2022}
Ko, H., Son, B., Lee, Y., Jang, H., Lee, J.: The economic value of nft: Evidence from a portfolio analysis using mean--variance framework. Finance Research Letters  \textbf{47},  102784 (2022)

\bibitem{kosba2016hawk}
Kosba, A., Miller, A., Shi, E., Wen, Z., Papamanthou, C.: Hawk: The blockchain model of cryptography and privacy-preserving smart contracts. In: IEEE symposium on security and privacy (SP). pp. 839--858. IEEE (2016)

\bibitem{kotzer2024sok}
Kotzer, A., Gandelman, D., Rottenstreich, O.: {SoK}: Applications of sketches and rollups in blockchain networks. IEEE Transactions on Network and Service Management (TNSM)  (2024)

\bibitem{kwon2019cosmos}
Kwon, J., Buchman, E.: Cosmos whitepaper. A Netw. Distrib. Ledgers  \textbf{27},  1--32 (2019)

\bibitem{kyriazis2020bitcoin}
Kyriazis, N.A.: Is bitcoin similar to gold? an integrated overview of empirical findings. Journal of Risk and Financial Management  \textbf{13}(5), ~88 (2020)

\bibitem{l2beat}
L2BEAT: {L2BEAT} - the state of the layer two ecosystem. \url{https://l2beat.com/scaling/risk}  (2024)

\bibitem{morgia_game_2023}
La~Morgia, M., Mei, A., Mongardini, A.M., Nemmi, E.N.: A game of {NFTs}: Characterizing {NFT} wash trading in the ethereum blockchain. In: IEEE International Conference on Distributed Computing Systems (ICDCS). pp. 13--24. IEEE (2023)

\bibitem{ledger}
Ledger: Ledger nano s plus. \url{https://shop.ledger.com/products/ledger-nano-s-plus}  (2024)

\bibitem{leitter_non-fungible_2023}
Leitter, Z., Cambria, E.: Non-fungible tokens: What makes them valuable? In: 2023 IEEE International Conference on Data Mining Workshops (ICDMW). pp. 750--756 (2023). \doi{10.1109/ICDMW60847.2023.00102}

\bibitem{li2023security}
Li, J.: On the security of optimistic blockchain mechanisms. Available at SSRN 4499357  (2023)

\bibitem{li2024bitcoin}
Li, N., Qi, M., et~al.: Bitcoin inscriptions: Foundations and beyond. IEEE International Conference on Blockchain and Cryptocurrency (ICBC)  (2024)

\bibitem{li2023transaction}
Li, R., Hu, X., et~al.: Transaction fairness in blockchains, revisited. Cryptology ePrint Archive  (2023)

\bibitem{li2022smart}
Li, R., et~al.: How do smart contracts benefit security protocols? arXiv preprint arXiv:2202.08699  (2022)

\bibitem{li2022sok}
Li, R., et~al.: {SoK}: Tee-assisted confidential smart contract. Proceedings on Privacy Enhancing Technologies (PETs)  \textbf{3},  1--21 (2022)

\bibitem{looksrare}
LooksRare: Looksrare website. \url{https://looksrare.org}  (2024)

\bibitem{luu2016making}
Luu, L., Chu, D.H., Olickel, H., Saxena, P., Hobor, A.: Making smart contracts smarter. In: Proceedings of the ACM SIGSAC Conference on Computer and Communications Security (CCS). pp. 254--269 (2016)

\bibitem{madoff1997pop}
Madoff, S.H.: Pop art: a critical history. Univ of California Press (1997)

\bibitem{magiceden}
MagicEden: Magic eden marketplace. \url{https://magiceden.io}  (2024)

\bibitem{male1982religious}
M{\^a}le, E.: Religious art from the twelfth to the eighteenth century. Princeton University Press (1982)

\bibitem{nftdigitalartinau}
Mallis, A.: {NFT} digital art gallery launches in australia. \url{https://www.digitalnationaus.com.au/news/nft-digital-art-gallery-launches-in-australia-579020}  (2024)

\bibitem{maouchi_understanding_2022}
Maouchi, Y., Charfeddine, L., El~Montasser, G.: Understanding digital bubbles amidst the {COVID}-19 pandemic: Evidence from {DeFi} and {NFTs}. Finance Research Letters  \textbf{47},  102584 (2022)

\bibitem{Marek2014}
Marek, P., Pavol, R.: {BIP}-44: Multi-account hierarchy for deterministic wallets (2014), bIP: 44, Standards Track \url{https://github.com/bitcoin/bips/blob/master/bip-0044.mediawiki}

\bibitem{Marek2013}
Marek, P., Pavol, R., Aaron, Voisine, S.B.: {BIP}-39: Mnemonic code for generating deterministic keys (2013), bIP: 39, Standards Track \url{https://github.com/bitcoin/bips/blob/master/bip-0039.mediawiki}

\bibitem{martin2018baroque}
Martin, J.R.: Baroque. Routledge (2018)

\bibitem{mayer2016ecdsa}
Mayer, H.: {ECDSA} security in bitcoin and ethereum: a research survey. CoinFaabrik  \textbf{28}(126), ~50 (2016)

\bibitem{quantum}
McCoy, K.: Quantum in sotheby's. \url{https://www.sothebys.com/en/buy/auction/2021/natively-digital-a-curated-nft-sale-2/quantum}  (2024)

\bibitem{mckelvey2015merchandising}
McKelvey, S., Sliffman, A.J.: The merchandising right gone awry: What moore can be said. Am. Bus. LJ  \textbf{52}, ~317 (2015)

\bibitem{matthias_2024}
Meg, M.: Renaissance. https://www.britannica.com/event/Renaissance  (2024)

\bibitem{mekacher2022rarity}
Mekacher, A., Bracci, A., Nadini, M., Martino, M., Alessandretti, L., Aiello, L.M., Baronchelli, A.: How rarity shapes the {NFT} market. arXiv preprint arXiv:2204.10243  \textbf{9} (2022)

\bibitem{metamask}
MetaMask: The ultimate crypto wallet for {DeFi}, web3 apps, and {NFTs} | {MetaMask}. \url{https://metamask.io}  (2024)

\bibitem{mingxiao2017review}
Mingxiao, D., Xiaofeng, M., Zhe, Z., Xiangwei, W., Qijun, C.: A review on consensus algorithm of blockchain. In: IEEE International Conference on Systems, Man, and Cybernetics (SMC). pp. 2567--2572. IEEE (2017)

\bibitem{nakamoto2008peer}
Nakamoto, S., Bitcoin, A.: A peer-to-peer electronic cash system. Bitcoin.--URL: https://bitcoin. org/bitcoin. pdf  \textbf{4}(2), ~15 (2008)

\bibitem{needham1993denial}
Needham, R.M.: Denial of service. In: Proceedings of the ACM Conference on Computer and Communications Security (CCS). pp. 151--153 (1993)

\bibitem{nees2002early}
Nees, L.: Early medieval art. Oxford University Press, USA (2002)

\bibitem{nelson2008patron}
Nelson, J.K., Zeckhauser, R.: The patron's payoff: conspicuous commissions in Italian Renaissance art. Princeton University Press (2008)

\bibitem{niu_unveiling_2024}
Niu, Y., Li, X., Peng, H., Li, W.: Unveiling wash trading in popular nft markets. In: Companion Proceedings of the ACM on Web Conference (WWW). pp. 730--733 (2024)

\bibitem{nugent2016improving}
Nugent, T., Upton, D., Cimpoesu, M.: Improving data transparency in clinical trials using blockchain smart contracts. F1000Research  \textbf{5} (2016)

\bibitem{opensea}
Opensea: Opensea website. \url{https://opensea.io}  (2024)

\bibitem{optimism}
Optimism: Optimism website. \url{https://www.optimism.io}  (2024)

\bibitem{othman_impact_2023}
Othman, M., Nasr, E., Nasr, J., Karam, L.: Impact of crypto art sentiment on art valuation. In: 2023 IEEE 4th International Multidisciplinary Conference on Engineering Technology (IMCET). pp. 259--263. IEEE (2023)

\bibitem{pandora}
{Pandora Labs LLC}: Pondora website. \url{https://www.pandora.build}  (2024)

\bibitem{panofsky1933classical}
Panofsky, E., Saxl, F.: Classical mythology in mediaeval art. Metropolitan Museum Studies  \textbf{4}(2),  228--280 (1933)

\bibitem{patreon}
Patreon: Patreon website. \url{https://www.patreon.com}  (2024)

\bibitem{paul2023digital}
Paul, C.: Digital art. Thames \& Hudson (2023)

\bibitem{penny1993materials}
Penny, N.: The materials of sculpture. Yale University Press (1993)

\bibitem{perez2021smart}
Perez, D., Livshits, B.: Smart contract vulnerabilities: Vulnerable does not imply exploited. In: USENIX Security Symposium (USENIX Security). pp. 1325--1341 (2021)

\bibitem{peters2016understanding}
Peters, G.W., Panayi, E.: Understanding modern banking ledgers through blockchain technologies: Future of transaction processing and smart contracts on the internet of money. Springer (2016)

\bibitem{pezoa2016foundations}
Pezoa, F., Reutter, J.L., Suarez, F., Ugarte, M., Vrgo{\v{c}}, D.: Foundations of {JSON} schema. In: Proceedings of the International Conference on World Wide Web (WWW). pp. 263--273 (2016)

\bibitem{pierro2019influence}
Pierro, G.A., Rocha, H.: The influence factors on ethereum transaction fees. In: IEEE/ACM International Workshop on Emerging Trends in Software Engineering for Blockchain (WETSEB). pp. 24--31. IEEE (2019)

\bibitem{pierro2022can}
Pierro, G.A., Tonelli, R.: Can solana be the solution to the blockchain scalability problem? In: IEEE International Conference on Software Analysis, Evolution and Reengineering (SANER). pp. 1219--1226. IEEE (2022)

\bibitem{Pieter2012}
Pieter, W.: {BIP}-32: Hierarchical deterministic wallets (2012), bIP: 32, Informational \url{https://github.com/bitcoin/bips/blob/master/bip-0032.mediawiki}

\bibitem{pinto-gutierrez_nft_2022}
Pinto-Guti{\'e}rrez, C., Gait{\'a}n, S., Jaramillo, D., Velasquez, S.: The {NFT} hype: What draws attention to non-fungible tokens? Mathematics  \textbf{10}(3), ~335 (2022)

\bibitem{pippin2002abstract}
Pippin, R.B.: What was abstract art?(from the point of view of hegel). Critical Inquiry  \textbf{29}(1),  1--24 (2002)

\bibitem{polygon}
{Polygon Labs}: Polygon website. \url{https://polygon.technology}  (2024)

\bibitem{prechtel2019evaluating}
Prechtel, D., Gro{\ss}, T., M{\"u}ller, T.: Evaluating spread of 'gasless send' in ethereum smart contracts. In: IFIP international conference on new technologies, mobility and security (NTMS). pp.~1--6. IEEE (2019)

\bibitem{preneel1994cryptographic}
Preneel, B.: Cryptographic hash functions. European Transactions on Telecommunications  \textbf{5}(4),  431--448 (1994)

\bibitem{solv}
Protocol, S.: Solv website. \url{https://solv.finance/}  (2024)

\bibitem{qin2021attacking}
Qin, K., Zhou, L., Livshits, B., Gervais, A.: Attacking the defi ecosystem with flash loans for fun and profit. In: International Conference on Financial Cryptography and Data Security (FC). pp. 3--32. Springer (2021)

\bibitem{reed2020decentralized}
Reed, D., Sporny, M., Longley, D., Allen, C., Grant, R., Sabadello, M., Holt, J.: Decentralized identifiers (dids) v1. 0. Draft Community Group Report  (2020)

\bibitem{reutter2001artists}
Reutter, M.A.: Artists, galleries and the market: Historical economic and legal aspects of artist-dealer relationships. Vill. Sports \& Ent. LJ  \textbf{8}, ~99 (2001)

\bibitem{roy2024unveiling}
Roy, S.S., Das, D., Bose, P., Kruegel, C., Vigna, G., Nilizadeh, S.: Unveiling the risks of {NFT} promotion scams. In: Proceedings of the International AAAI Conference on Web and Social Media. vol.~18, pp. 1367--1380 (2024)

\bibitem{safe}
Safe: Safe wallet. \url{https://app.safe.global/}  (2024)

\bibitem{sanka2021systematic}
Sanka, A.I., Cheung, R.C.: A systematic review of blockchain scalability: Issues, solutions, analysis and future research. Journal of Network and Computer Applications  \textbf{195},  103232 (2021)

\bibitem{ledger2018}
SAS, L.: Ledger live : Most trusted \& secure crypto wallet. \url{https://metamask.io}  (2024)

\bibitem{scoville2007cosmic}
Scoville, N., Aussel, H., Brusa, M., Capak, P., Carollo, C.M., Elvis, M., Giavalisco, M., Guzzo, L., Hasinger, G., Impey, C., et~al.: The cosmic evolution survey ({COSMOS}): overview. The Astrophysical Journal Supplement Series  \textbf{172}(1), ~1 (2007)

\bibitem{shirole2020cryptocurrency}
Shirole, M., Darisi, M., Bhirud, S.: Cryptocurrency token: An overview. In: Proceedings of the International Conference on Blockchain Technology (IC-BCT). pp. 133--140. Springer (2020)

\bibitem{sifat_suspicious_2024}
Sifat, I., Tariq, S.A., van Donselaar, D.: Suspicious trading in nonfungible tokens ({NFTs}). Information \& Management  \textbf{61}(1),  103898 (2024)

\bibitem{skillshare}
Skillshare: Skillshare website. \url{https://www.skillshare.com}  (2024)

\bibitem{smith2009contemporary}
Smith, T.E.: What is contemporary art? University of Chicago Press (2009)

\bibitem{sotheby}
Sotheby: Sotheby website. \url{https://www.sothebys.com}  (2024)

\bibitem{sudoswap}
Sudoswap: Sudoswap website. \url{https://sudoswap.xyz}  (2024)

\bibitem{defihacklabs}
SunWeb3Sec: Defi hacks reproduce - foundry. \url{https://github.com/SunWeb3Sec/DeFiHackLabs}  (2024)

\bibitem{superrare}
Superrare: Superrare website. \url{https://superrare.co}  (2024)

\bibitem{suratkar2020cryptocurrency}
Suratkar, S., Shirole, M., Bhirud, S.: Cryptocurrency wallet: A review. In: International Conference on Computer, Communication and Signal Processing (ICCCSP). pp.~1--7. IEEE (2020)

\bibitem{Nick1997}
Szabo, N.: The idea of smart contracts. Nick Szabo's Papers and Concise Tutorials  \textbf{6} (1997)

\bibitem{tahmasbi_identifying_2024}
Tahmasbi, N., Shan, G., French, A.M.: Identifying washtrading cases in nft sales networks. IEEE Transactions on Computational Social Systems (TCSS)  (2023)

\bibitem{tang2021analysis}
Tang, Y., Xu, C., Zhang, C., Wu, Y., Zhu, L.: Analysis of address linkability in tornado cash on {E}thereum. In: China Cyber Security Annual Conference. pp. 39--50. Springer (2021)

\bibitem{telegram}
Telegram: Telegram website. \url{https://telegram.org}  (2024)

\bibitem{thompson2011art}
Thompson, D.: Art fairs: The market as medium. Negotiating Values in the Creative Industries: Fairs, Festivals and Competitive Events pp. 59--72 (2011)

\bibitem{thuy2020fast}
Thuy, N.T.T., Khai, L.D., et~al.: A fast approach for bitcoin blockchain cryptocurrency mining system. Integration  \textbf{74},  107--114 (2020)

\bibitem{tikTok}
TikTok: Tiktok website. \url{https://www.tiktok.com}  (2024)

\bibitem{tolmach2021survey}
Tolmach, P., Li, Y., Lin, S.W., Liu, Y., Li, Z.: A survey of smart contract formal specification and verification. ACM Computing Surveys (CSUR)  \textbf{54}(7),  1--38 (2021)

\bibitem{tosic_beyond_2023}
Tošić, A., Hrovatin, N., Vičič, J.: Beyond the surface: Advanced wash trading detection in decentralized {NFT} markets, \url{http://arxiv.org/abs/2312.16603}

\bibitem{trezor}
Trezor: Trezor. \url{https://www.fxempire.com/crypto/wallets/trezor}  (2024)

\bibitem{tron}
TRON: Tron network. \url{https://tron.network}  (2024)

\bibitem{udemy}
Udemy: Udemy website. \url{https://www.udemy.com}  (2024)

\bibitem{urom_dynamic_2022}
Urom, C., Ndubuisi, G., Guesmi, K.: Dynamic dependence and predictability between volume and return of non-fungible tokens ({NFTs}): The roles of market factors and geopolitical risks. Finance Research Letters  \textbf{50},  103188 (2022)

\bibitem{von_wachter_nft_2022}
Von~Wachter, V., Jensen, J.R., Regner, F., Ross, O.: Nft wash trading: Quantifying suspicious behaviour in nft markets. In: International Conference on Financial Cryptography and Data Security. pp. 299--311. Springer (2022)

\bibitem{atomicwallet}
Wallet, A.: Atomic wallet. \url{https://atomicwallet.io}  (2024)

\bibitem{trustwallet}
Wallet, T.: Trust wallet. \url{https://trustwallet.com}  (2024)

\bibitem{wan2021smart}
Wan, Z., Xia, X., Lo, D., Chen, J., Luo, X., Yang, X.: Smart contract security: A practitioners' perspective. In: IEEE/ACM 43rd International Conference on Software Engineering (ICSE). pp. 1410--1422. IEEE (2021)

\bibitem{wang2023exploring}
Wang, G., et~al.: Exploring blockchains interoperability: A systematic survey. ACM Computing Surveys  \textbf{55}(13s),  1--38 (2023)

\bibitem{wang2021non}
Wang, Q., Li, R., Wang, Q., Chen, S.: Non-fungible token ({NFT}): Overview, evaluation, opportunities and challenges. arXiv preprint arXiv:2105.07447  (2021)

\bibitem{wang2022exploring}
Wang, Q., Li, R., Wang, Q., Chen, S., Ryan, M., Hardjono, T.: Exploring web3 from the view of blockchain. arXiv preprint arXiv:2206.08821  (2022)

\bibitem{wang2020preserving}
Wang, Q., Qin, B., Hu, J., Xiao, F.: Preserving transaction privacy in bitcoin. Future Generation Computer Systems  \textbf{107},  793--804 (2020)

\bibitem{wang2023understanding}
Wang, Q., Yu, G.: Understanding {BRC}-20: Hope or hype. Available at SSRN 4590451  (2023)

\bibitem{wang2024bridging}
Wang, Q., Yu, G., Chen, S.: Bridging {BRC}-20 to {E}thereum. IEEE International Conference on Blockchain and Cryptocurrency (ICBC) pp. 1--10 (2024)

\bibitem{wang2024cryptocurrency}
Wang, Q., Yu, G., Chen, S.: Cryptocurrency in the aftermath: Unveiling the impact of the {SVB} collapse. IEEE Transactions on Computational Social Systems (TCSS)  (2024)

\bibitem{Saber2022}
Wang, Q., Yu, G., Fu, S., Chen, S., Yu, J., Xu, X.: A referable {NFT} scheme. In: IEEE International Conference on Blockchain and Cryptocurrency (ICBC). pp.~1--6. IEEE (2023)

\bibitem{wang2019blockchain}
Wang, S., Ouyang, L., Yuan, Y., Ni, X., Han, X., Wang, F.Y.: Blockchain-enabled smart contracts: architecture, applications, and future trends. IEEE Transactions on Systems, Man, and Cybernetics: Systems  \textbf{49}(11),  2266--2277 (2019)

\bibitem{wang2022volatility}
Wang, Y.: Volatility spillovers across {NFTs} news attention and financial markets. International Review of Financial Analysis  \textbf{83},  102313 (2022)

\bibitem{waysdorf2021remix}
Waysdorf, A.S.: Remix in the age of ubiquitous remix. Convergence  \textbf{27}(4),  1129--1144 (2021)

\bibitem{weisstein1967expressionism}
Weisstein, U.: Expressionism: Style or" weltanschauung"? Criticism  \textbf{9}(1),  42--62 (1967)

\bibitem{wen_nftdisk_2023}
Wen, X., Wang, Y., Yue, X., Zhu, F., Zhu, M.: {NFTDisk}: Visual detection of wash trading in {NFT} markets. In: Proceedings of the CHI Conference on Human Factors in Computing Systems. pp. 1--15 (2023)

\bibitem{werner2022sok}
Werner, S., Perez, D., Gudgeon, L., Klages-Mundt, A., Harz, D., Knottenbelt, W.: {SoK}: Decentralized finance (defi). In: Proceedings of the ACM Conference on Advances in Financial Technologies. pp. 30--46 (2022)

\bibitem{white2022characterizing}
White, B., Mahanti, A., Passi, K.: Characterizing the opensea {NFT} marketplace. In: Companion Proceedings of the Web Conference (WWW). pp. 488--496 (2022)

\bibitem{Wang2020}
Will, W., Mike, M., Yi, C., Ryan, C., Zhongxin, W.: {ERC}-3525: Semi-fungible token (2020), \url{https://eips.ethereum.org/EIPS/eip-3525}

\bibitem{William2018}
William, E., Dieter, S., Jacob, E., Nastassia, S.: {ERC}-721: Non-fungible token standard (2018), ethereum Improvement Protocol, EIP-721 \url{https://eips.eth ereum.org/EIPS/eip-721}

\bibitem{winckelmann1880history}
Winckelmann, J.J.: The history of ancient art, vol.~2. Boston: JR Osgood and Company (1880)

\bibitem{Witek2018}
Witek, R., Andrew, C., Philippe, C., James, T., Eric, B., Ronan, S.: {ERC}-1155: Multi token standard (2018), ethereum Improvement Protocol, EIP-1155 \url{https://eips.ethereum.org/EIPS/eip-1155}

\bibitem{wood2016polkadot}
Wood, G.: Polkadot: Vision for a heterogeneous multi-chain framework. White paper  \textbf{21}(2327), ~4662 (2016)

\bibitem{wood2014ethereum}
Wood, G., et~al.: Ethereum: A secure decentralised generalised transaction ledger. Ethereum project yellow paper  \textbf{151}(2014),  1--32 (2014)

\bibitem{x}
X: X website. \url{https://x.com}  (2024)

\bibitem{x2y2}
X2Y2: X2y2 marketplace. \url{https://x2y2.io}  (2024)

\bibitem{xu2019anatomy}
Xu, J., Livshits, B.: The anatomy of a cryptocurrency $\{$Pump-and-Dump$\}$ scheme. In: {USENIX} Security Symposium ({USENIX} Security). pp. 1609--1625 (2019)

\bibitem{xu2023sok}
Xu, J., Paruch, K., Cousaert, S., Feng, Y.: Sok: Decentralized exchanges (dex) with automated market maker (amm) protocols. ACM Computing Surveys  \textbf{55}(11),  1--50 (2023)

\bibitem{xu2018blendcac}
Xu, R., Chen, Y., Blasch, E., Chen, G.: Blendcac: A smart contract enabled decentralized capability-based access control mechanism for the iot. Computers  \textbf{7}(3), ~39 (2018)

\bibitem{yakovenko2018solana}
Yakovenko, A.: Solana: A new architecture for a high performance blockchain v0. 8.13. Whitepaper  (2018)

\bibitem{yang2023non}
Yang, J., Li, Y., Lai, Y., Liu, M.: Non-fungible tokens ({NFTs}): tokens of digital assets on the blockchain. In: Proceedings of the International Conference on Electronics, Computers and Communication Technology. pp. 175--182 (2023)

\bibitem{yang2023role}
Yang, W., Shi, Z.J., Lin, S.: The role of royalty fee in nft markets. Song, The Role of Royalty Fee in NFT Markets (May 5, 2023)  (2023)

\bibitem{yousaf_relationship_2022}
Yousaf, I., Yarovaya, L.: The relationship between trading volume, volatility and returns of non-fungible tokens: evidence from a quantile approach. Finance Research Letters  \textbf{50},  103175 (2022)

\bibitem{yu2024toward}
Yu, G., Wang, X., et~al.: Toward web3 applications: Easing the access and transition. IEEE Transactions on Computational Social Systems  (2024)

\bibitem{yu2024maximizing}
Yu, G., et~al.: Maximizing nft incentives: References make you rich. arXiv preprint arXiv:2402.06459  (2024)

\bibitem{boredapeyachtclub}
YUGALABS: Bored ape yacht club. \url{https://boredapeyachtclub.com}  (2024)

\bibitem{Zach2018}
Zach, B., James, M., Blaine, M., James, S.: {ERC}-2981: {NFT} royalty standard (2018), ethereum Improvement Protocol, EIP-2981 \url{https://eips.ethereum.org/EIPS/eip-2981}

\bibitem{zelenyanszki2023linking}
Zelenyanszki, D., H{\'o}u, Z., Biswas, K., Muthukkumarasamy, V.: Linking nft transaction events to identify privacy risks. In: International Symposium on Distributed Ledger Technology. pp. 82--97. Springer (2023)

\bibitem{zetzsche2020decentralized}
Zetzsche, D.A., Arner, D.W., Buckley, R.P.: Decentralized finance (defi). Journal of Financial Regulation  \textbf{6},  172--203 (2020)

\bibitem{zheng2020overview}
Zheng, Z., Xie, S., Dai, H.N., Chen, W., Chen, X., Weng, J., Imran, M.: An overview on smart contracts: Challenges, advances and platforms. Future Generation Computer Systems  \textbf{105},  475--491 (2020)

\bibitem{zheng2017overview}
Zheng, Z., Xie, S., Dai, H., Chen, X., Wang, H.: An overview of blockchain technology: Architecture, consensus, and future trends. In: IEEE International Congress on Big Data (BigData Congress). pp. 557--564. Ieee (2017)

\bibitem{zhou2020solutions}
Zhou, Q., Huang, H., Zheng, Z., Bian, J.: Solutions to scalability of blockchain: A survey. Ieee Access  \textbf{8},  16440--16455 (2020)

\bibitem{ziemke_what_2023}
Ziemke, V., Estermann, B., Wattenhofer, R., Wang, Y.: What determines the price of {NFTs}? In: 2023 IEEE 29th International Conference on Parallel and Distributed Systems (ICPADS). pp. 1562--1568. IEEE (2023)

\end{thebibliography}
\end{document}